\RequirePackage{fixltx2e}

\documentclass[aps,prb,12pt,onecolumn,
               amsmath,amssymb,amsfonts,
               notitlepage]{revtex4-1}

\usepackage[varg]{txfonts}
\usepackage{hyperref}
\usepackage{amsmath}
\usepackage{color}
\usepackage{graphicx}
\usepackage[percent]{overpic}
\usepackage{caption}
\usepackage{subcaption}
\usepackage{bm}
\usepackage{pict2e}

\newcommand{\ket}[1]{\left| #1  \right \rangle}

\newcommand{\avg}[1]{\langle #1 \rangle}

\newcommand{\h}[1]{{#1}^{\dagger}}

\newcommand{\up}{\uparrow}
\newcommand{\down}{\downarrow}

\newcommand{\tsup}[1]{\textsuperscript{#1}}
\newcommand{\tsub}[1]{\textsubscript{#1}}

\newcommand{\eV}{\ {\rm eV}}
\newcommand{\meV}{\ {\rm meV}}

\newcommand{\K}{\ {\rm K}}

\newcommand{\hc}{{\rm h.c.}}

\newcommand{\ir}{Ir\tsup{4+}}
\newcommand{\nairo}{Na\tsub{2}IrO\tsub{3}}
\newcommand{\naliiro}{(Na\tsub{1-x}Li\tsub{x})\tsub{2}IrO\tsub{3}}
\newcommand{\nairtio}{Na\tsub{2}(Ir\tsub{{1-x}}Ti\tsub{x})O\tsub{3}}
\newcommand{\liirtio}{Li\tsub{2}(Ir\tsub{{1-x}}Ti\tsub{x})O\tsub{3}}

\newcommand{\anairo}{\nairo{}}
\newcommand{\hknairo}{Na\tsub{4}Ir\tsub{3}O\tsub{8}}

\newcommand{\aliiro}{$\alpha$-Li\tsub{2}IrO\tsub{3}}
\newcommand{\bliiro}{$\beta$-Li\tsub{2}IrO\tsub{3}}
\newcommand{\gliiro}{$\gamma$-Li\tsub{2}IrO\tsub{3}}
\newcommand{\srruo}{Sr\tsub{2}RuO\tsub{4}}
\newcommand{\sriro}{Sr\tsub{2}IrO\tsub{4}}
\newcommand{\srirob}{Sr\tsub{3}Ir\tsub{2}O\tsub{7}}
\newcommand{\lacuo}{La\tsub{2}CuO\tsub{4}}

\newcommand{\srirop}{SrIrO\tsub{3}}

\newcommand{\srtio}{SrTiO\tsub{3}}

\newcommand{\arucl}{$\alpha$-RuCl\tsub{3}}
\newcommand{\rps}{Sr\tsub{n+1}Ir\tsub{n}O\tsub{3n+1}}

\renewcommand{\vec}[1]{\bm{\mathbf{#1}}}

\newcommand{\vhat}[1]{\vec{\hat{#1}}}
\newcommand{\pS}{J}
\newcommand{\muB}{\mu_{\rm B}}

\definecolor{cred}{RGB}{228,26,28}
\definecolor{cblue}{RGB}{55,126,184}
\definecolor{cgreen}{RGB}{77,175,74}
\definecolor{cgray}{RGB}{150,150,150}
\definecolor{clgray}{RGB}{200,200,200}
\definecolor{cpurple}{RGB}{152,78,163}
\definecolor{corange}{RGB}{255,127,0}
\definecolor{cgold}{RGB}{230,171,2}

\hypersetup{colorlinks=true,linkcolor=cblue,citecolor=cblue,urlcolor=cblue} 

\begin{document}

\title{Spin-Orbit Physics Giving Rise to Novel Phases in Correlated Systems; Iridates and Related Materials}

\author{Jeffrey G. Rau}
\affiliation{
Department of Physics and Astronomy, University of Waterloo, Ontario,
N2L 3G1, Canada
}

\author{Eric Kin-Ho Lee}
\affiliation{
Department of Physics and Center for Quantum Materials, University of
Toronto, Toronto, Ontario, M5S 1A7, Canada
}

\author{Hae-Young Kee}
\email[Electronic Address: ]{hykee@physics.utoronto.ca}
\affiliation{
Department of Physics and Center for Quantum Materials, University of
Toronto, Toronto, Ontario, M5S 1A7, Canada
}
\affiliation{
Canadian Institute for Advanced Research/Quantum Materials Program,
Toronto, Ontario, MSG 1Z8, Canada
}

\begin{abstract}
Recently, the effects of spin-orbit coupling (SOC) in correlated materials
have become one of the most actively studied subjects in condensed
matter physics, as correlations and SOC together can lead to the
discovery of new phases. Among candidate materials, iridium oxides
(iridates) have been an excellent playground to uncover such novel
phenomena.  In this review, we discuss recent progress in iridates and
related materials, focusing on the basic concepts, relevant
microscopic Hamiltonians, and unusual properties of iridates in
perovskite- and honeycomb-based structures. Perspectives on SOC and
correlation physics beyond iridates are also discussed.
\end{abstract}

\date{\today}

\maketitle

\vfill


\clearpage

\tableofcontents

\section{Introduction}

Spin-orbit coupling (SOC) is a relativistic effect that links the
orbital and spin angular momenta of an electron. Though suppressed by
the fine structure constant, the electric fields near the nuclei of
atoms with a large number of protons can render this interaction
significant. The natural place to find significant SOC is thus in
atoms with high atomic numbers, moving down the rows of the periodic
table into the heavier elements.

The effects of SOC in materials with such heavy atoms have been
studied intensively in the context of semi-conductors. In these weakly
correlated materials, SOC entangles the crystal momentum and spin of
the electron, locking the kinetic and internal degrees of freedom
together. This leads to a number of intriguing phenomena, particularly
in transport; examples include the anomalous Hall effect (AHE) and the
control of spin currents being applied in the field of spintronics.
More recently it has been found that SOC plays an essential role in
the fast growing field of topological insulators (TI) and
metals. Many, if not all, of the experimental examples of such
topological phases have been found in these types of heavy
semi-conductors.

While the effects of SOC in weakly correlated materials described
above has been thoroughly considered, its importance in more strongly
correlated transition metal materials remains less developed.  There
are several energy scales to consider in such materials: the atomic
interactions (schematically) on-site Hubbard interaction $U$, Hund's
coupling $J_H$, the SOC $\lambda$, the crystal field $\Delta$ and the
electron kinetic energy described by hopping integral $t$.  Much
theoretical and experimental effort has been brought to bear on $3d$
transition metals such as high temperature cuprates, manganites, and
vanadium oxides. In these compounds the atomic interactions, crystal
field and kinetic terms can generally compete, though the SOC remains
small.  In heavy transition metals such as those with $5d$ and even
$4d$ electrons SOC is significant as well, and so all of these energy
scales can be comparable. As one moves through the different heavy
transition metal materials, small changes in these details can tip the
scales, revealing a surprisingly rich family of behaviours.  This can
be contrasted with $4f$ or $5f$ electrons in the lanthanides or
actinides, where the electron interactions are dominant, followed by
the SOC and then the crystal fields.

While intensive studies have recently been undertaken, we are still
far from complete understanding on the combined effects of SOC and
electronic correlation. Material examples of $4d$ and $5d$ systems can
be found for a variety of two- and three-dimensional lattices with
varying degrees of frustration, covering the range from weakly to
strongly correlated. An excellent overview of this physics is provided
in a recent review \cite{krempa2014correlated}, with a focus on
examples from the pyrochlore iridates and the $5d$ double-perovskite
magnets.  In this review we will focus on magnetism, topological phases and
superconductivity induced by the competition between strong SOC
and electronic correlation.  We illustrate these concepts with
examples drawn from the perovskite and honeycomb iridium oxides. After
a basic introduction to the relevant models in
Sec. \ref{sec:background}, we begin with perovskite iridates in
Sec. \ref{sec:perovskites}. Drawing analogies to the related cuprates
and ruthenates, we discuss proposals for realizing topological and
superconducting phases in the materials. In Sec. \ref{sec:kitaev} we
tackle the burgeoning field of Kitaev magnetism. In particular, we
offer a summary of the current status of two- and three-dimensional
honeycomb iridates. Concluding remarks and outlook are provided in
Sec. \ref{sec:outlook}.

\section{Background}
\label{sec:background}
\begin{figure}[tp]
    \begin{overpic}[width=0.8\textwidth]
    {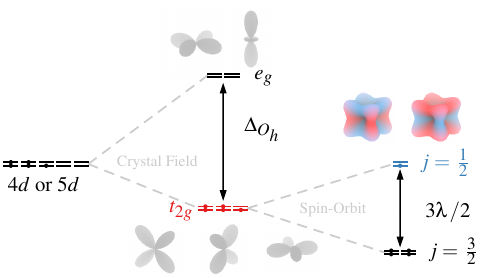}
    \put(0,30){\includegraphics[width=3cm]{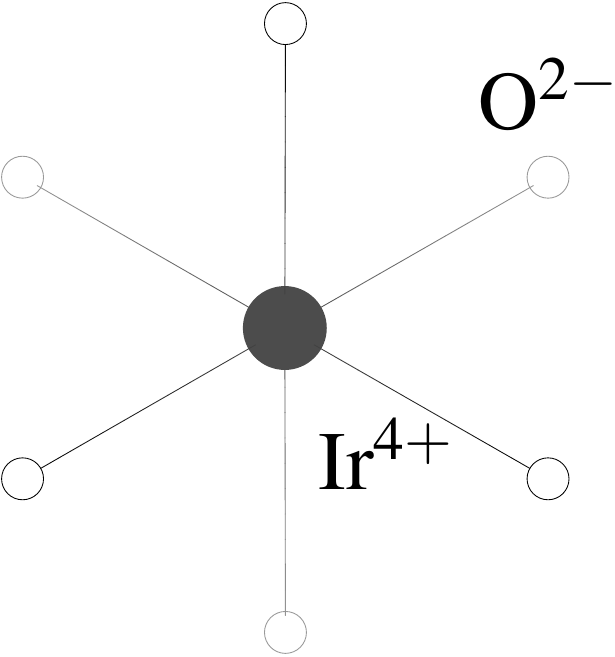}}
    \put(0,52){(a)}
    \put(0,27){(b)}
    \put(65,33){(c)}
  \end{overpic}
  \caption{
    \label{fig:levels}
    (a) Octahedral geometry of transition metal site illustrated for
an iridium oxide.  (b) Splitting of the $4d$ or $5d$ levels by
octahedral crystal fields $\Delta_{O_h}$ and by SOC
$\lambda$ into $j=1/2$ and $j=3/2$ levels. (c) Illustration
of the atomic $j=1/2$ wave-functions. The composition of the
$j=1/2$ states with the spin-$\up$ charge density
is shown in red and the spin-$\down$ charge density in blue.
  }
\end{figure}

The building blocks of our discussion are the atomic states of the
partially filled $4d$- or $5d$ ions. In solids of interest, these
states are split by a predominantly octahedral crystal field potential
into a $t_{2g}$ triplet and $e_g$ doublet, as shown in
Fig. \ref{fig:levels}. The energy gap to the $e_g$ doublet is large,
so these states can be safely ignored when we consider electron
filling less than six.  We are primarily interested in the $d^5$
configuration, which can be regarded as single hole in one of the
$t_{2g}$ states. When projected into this manifold, the angular
momentum of the the $d$ electrons is mapped to a set of effective
$l=1$ angular momentum operators, $-\vec{L}$. The large SOC then acts
within the $t_{2g}$ manifold as $-\lambda \vec{L} \cdot \vec{S}$ where
$\vec{L}$ is an effective $l=1$ angular momentum and $\vec{S}$ is the
spin.  Using the rules of addition of angular momenta we see that SOC
splits the $t_{2g}$ multiplet into an effective $j=1/2$ doublet and
effective $j=3/2$ quartet as show in Fig. \ref{fig:levels}. The
$j=3/2$ states are lower in energy and separated from the $j=1/2$
states by a gap of $3\lambda/2$.  Written in terms of the $t_{2g}$
states one has
\begin{subequations}
\label{jeff}
\begin{eqnarray}
  \ket{\frac{1}{2},\pm\frac{1}{2}} &=& \sqrt{\frac{1}{3}}\left(
                                       \ket{yz,\mp} \pm i\ket{xz,\mp} 
                                       \pm \ket{xy,\pm}
                                       \right), \\
  \ket{\frac{3}{2},\pm\frac{3}{2}} &=& \sqrt{\frac{1}{2}}\left(
                                       \ket{yz,\pm} \pm i\ket{xz,\pm}\right), \\
  \ket{\frac{3}{2},\pm\frac{1}{2}} &=& \sqrt{\frac{1}{6}}\left(
                                       \ket{yz,\mp} \pm i\ket{xz,\mp} 
                                       - 2\ket{xy,\pm}
                                       \right),
\end{eqnarray}
\end{subequations}
where $\ket{yz,\pm},\ket{xz,\pm},\ket{xy,\pm}$ are the $t_{2g}$ states
and spin up and down correspond to $\pm$. The important role played by
SOC can be seen in the entanglement of the spin and orbital degrees of
freedom in these wave-functions, as illustrated in
Fig. \ref{fig:levels}. The states of this $j=1/2$ doublet and its
pseudo-spin operators $\vec{\pS}$ will be a common thread in our
discussion of the family of iridium oxides and ruthenium materials.
While strongly spin-orbit entangled, it remains \emph{isotropic}, with
a $g$-factor of $-2$. Explicitly, the magnetic moment operator
$\vec{\mu} = \muB \left(\vec{L} + 2\vec{S}\right)$ becomes $-2 \muB
\vec{J}$ when projected into the $j=1/2$ states. The anisotropy of
these systems thus manifests predominantly in the \emph{coupling} of
$j=1/2$ moments \cite{khaliullin2005orbital}, not in the single-ion
properties.

The interactions of $j=1/2$ electrons are determined by the atomic
interactions of the free ion projected into the $t_{2g}$ manifold and
the kinetic terms, or hoppings of the $t_{2g}$
electrons. Schematically we can write the general multi-orbital model
\begin{equation}
  \label{eq:full-model}
  \sum_{ij} \sum_{\alpha\beta} \sum_{\sigma} 
  t^{\alpha\beta}_{ij}\left(
    \h{d}_{i\alpha\sigma} d_{j\beta\sigma} +{\rm h.c}
  \right) + 
  \sum_i \left[
   \left(\frac{U-3J_H}{2}\right)\left(N_i-5\right)^2
   - 2J_H \vec{S}^2_i - \frac{J_H}{2} \vec{L}_i^2 
   - \lambda \vec{L}_i \cdot \vec{S}_i
 \right],
\end{equation}
where $N_i$ is the total number operator, $\vec{S}_i$ is the total
spin operator and $\vec{L}_i$ is the total pseudo-angular momentum
operator at site $\vec{r}_i$. The $\h{d}_{i\alpha\sigma}$ operator
creates a $t_{2g}$ electron in orbital $\alpha=yz,xz$ or $xy$ with
spin $\sigma = \up,\down$ at site $\vec{r}_i$ and
$t^{\alpha\beta}_{ij}$ are the tight-binding hopping parameters.  The
hopping amplitudes $t^{\alpha\beta}_{ij}$ will have contributions
coming from direct $d$-$d$ overlap, as well as from processes through
the intermediate oxygen atoms. As these kinetic terms are strongly
material dependent in both structure and scale, we save discussion of
their details for the sections devoted to each specific class of
material.

The local atomic interactions are more generic and can be expressed in
terms of two parameters \cite{georges2013strong}: the Coulomb
repulsion $U$ and the Hund's coupling $J_H$.  We have added a chemical
potential to favour the $N=5$ state relevant for Ir$^{4+}$ and
Ru$^{3+}$ and have fixed the inter-orbital repulsion $U'$ to the free
ion value $U-2J_H$. Typically one expects $U \sim 2\eV$
\cite{van1988electron}, $J_H \sim 0.2\eV$ and $3\lambda/2 \sim
0.4-0.5\eV$ \cite{kim2008novel} for an \ir{} ion.  On the other hand
for a Ru$^{3+}$ ion, $U$ and $J_H$ are stronger while atomic SOC is
weaker than the Ir\tsup{4+} case.  Due to screening effects in a
solid, these parameters will be renormalized, and the free ion
relation $U'=U-2J_H$ can be violated. Typically, one expects the
Coulomb integral $U$ to be more strongly screened than the Hund's
coupling $J_H$ \cite{van1988electron,georges2013strong}.

There are several limiting regimes where the physics of this model is
particularly clear. In Sec. \ref{sec:perovskites} we will consider the
limit where the interactions between the $j=1/2$ and $j=3/2$ electrons
can be neglected. This yields an effective single-band Hubbard model
for the $j=1/2$ electrons.  With sufficiently weak correlations, this
approximation can be useful as a starting point for itinerant single
band $j=1/2$ systems. In the strongly correlated regime, the $j=1/2$
electrons localize and one is left with a pseudo-spin model.  By
projecting into only the $j=1/2$ states one loses the $j=3/2$ excited
states and the effects of Hund's coupling. As we will see in
Sec. \ref{sec:kitaev}, for the honeycomb iridates, this leading term
may \emph{cancel}, so the contribution of the virtual processes that go
through $j=3/2$ states become the largest interactions.

\section{Magnetism, topological phases and superconductivity in perovskite iridates}
\label{sec:perovskites}
\begin{figure}
\begin{subfigure}[b]{0.2\textwidth}
  \begin{overpic}[height=3\textwidth]
    {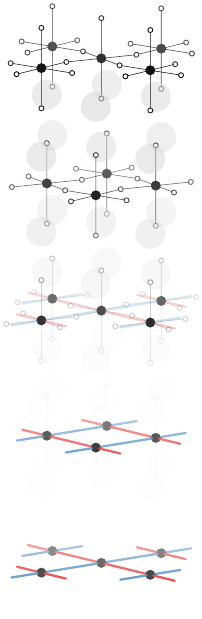}
    \put(-0.5,77){\textcolor{cgray}{Sr\tsup{2+}}}
    \put(13,45){Ir\tsup{4+}}
    \put(30,72){O\tsup{2-}}
    \linethickness{0.75pt}
    \put(2,15){\color{black}\vector(0,1){8}}
    \put(2,15){\color{black}\vector(1,0){8}}
    \put(2,15){\color{black}\vector(1,2){2}}
    \put(1.5,23.5){$\scriptsize \vhat{c}$}
    \put(4,19){$\scriptsize \vhat{b}$}
    \put(10.5,14.5){$\scriptsize \vhat{a}$}
  \end{overpic}
  \caption{\sriro{} $(n=1)$}
\end{subfigure} \hspace{0.2cm}
\begin{subfigure}[b]{0.2\textwidth}
  \includegraphics[height=3\textwidth]{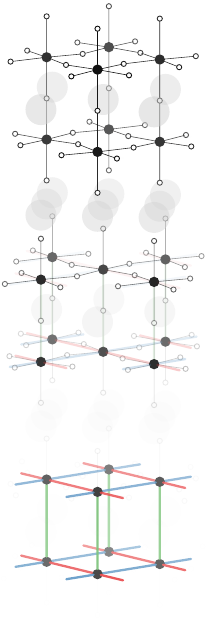}
  \caption{\srirob{} $(n=2)$}
\end{subfigure} \hspace{0.2cm}
\begin{subfigure}[b]{0.2\textwidth}
  \includegraphics[height=3\textwidth]{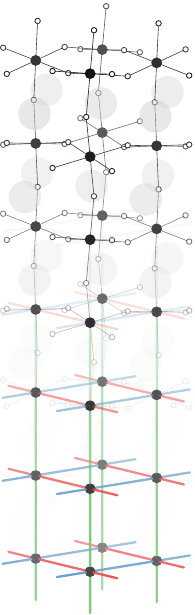}
  \caption{\srirop{} $(n=\infty)$}
\end{subfigure}
\caption{\label{fig:ruddlesden}
  Crystal structures of the perovskite iridates from the
Ruddlesden-Popper series \rps{}. These include the quasi-two
dimensional (a) single-layer \sriro{}
\cite{crawford1994structural,huang1994neutron}.  and (b) bilayer
\srirob{} \cite{cao2002anomalous} and the three dimensional limit (c)
orthorhombic \srirop{} \cite{longo1971structure,zhao2008high}.
}
\end{figure}

Historically, the field began with the study of \sriro{} and \srruo{}
as isostructural analogues of \lacuo{}, the parent compound of the
cuprate superconductors. While \sriro{} was first synthesized in 1957
\cite{randall1957preparation}, early work focused mainly on chemical
aspects, with variety of ternary and quaternary oxides of iridium and
ruthenium reported in the following decades. \cite{komer1978ternary}
More serious interest in \sriro{} was piqued when superconductivity
was discovered in \srruo{} \cite{maeno1994superconductivity}, but an
insulating state with small ferromagnetic moment was found instead
\cite{crawford1994structural, cao1998weak}.  Interest was renewed by
the discovery that the interplay between SOC and electronic
correlations were driving the physics in this material
\cite{kim2008novel,kim2009phase}.

In this section, our primary focus will be on magnetism in \sriro{}
and \srirob{} and the possible realization of novel topological phases
and superconductivity. We start with a review of the experimental and
theoretical progress in understanding perovskite iridates such as the
Ruddlesden-Popper series of \rps{} through a discussion of the common
building block of corner-shared octahedra in both the itinerant and
localized limits. Next, we discuss some candidate topological phases
built from iridium oxide heterostructures. Finally, we discuss the
effects of charge doping, with an eye toward the possibility of
superconductivity in \sriro{}.

\subsection{Magnetism}
\label{sec:magnetism}
In the single and bilayer perovskites \sriro{} and \srirob{}
\cite{kafalas1972high} evidence from the optical gap
\cite{moon2008dimensionality} to the magnetic ordering transition
\cite{cao1998weak,nagai2007canted,kim2012dimensionality,fujiyama2012weak}
have indicated that both are Mott insulators, though \srirob{} being
substantially weaker than \sriro{}.  The modest on-site Hubbard
interaction $U$ proves sufficient to localize the electrons due to the
narrow, spin-orbit coupled $j=1/2$ bands.  In \sriro{} the magnetic
ordering is an in-plane, canted antiferromagnet
\cite{kim2009phase,ye2013magnetic} following the staggered rotation of
the oxygen octahedra, as illustrated in Fig. \ref{fig:rotations}.
Though the net ferromagnetic moment of this state was observed in
early studies \cite{crawford1994structural,
huang1994neutron,cao1998weak}, the crucial role played by strong SOC
was only understood fairly recently \cite{jackeli2009mott}.  The
moment directions in \srirob{} are very different, forming a collinear
arrangement perpendicular to the iridium planes
\cite{fujiyama2012weak} as shown in Fig. \ref{fig:order-bilayer}.

The bulk trilayer and higher compounds ($n \geq 3$) are unstable
under ambient pressure \cite{matsuno2014engineering}, including the
three-dimensional limit \srirop{} which takes a post-perovskite structure 
\cite{longo1971structure}. An orthorhombic perovskite \srirop{} can
be stabilized via the application of pressure;
this is a semi-metal and shows no evidence of magnetic ordering
\cite{zhao2008high,moon2008dimensionality}. One can then posit
\cite{moon2008dimensionality} a metal-insulator transition (MIT) exists as a
function of layer count near $n_c \sim 3$, placing \sriro{} ($n=1$)
and \srirob{} ($n=2$) on the insulating side and orthorhombic \srirop{}
($n=\infty$) on the metallic side.  Due to the proximity to this
putative MIT there has been some debate on the
applicability of the strongly coupled Mott picture as compared to a
more weakly coupled Slater picture
\cite{arita2012ab,hsieh2012observation,carter2013theory}.  Since the
weak and strong-coupling regimes give the same magnetic orderings
,\cite{carter2013theory} this distinction may be somewhat academic at
low temperature. Near the transition the Mott and Slater regimes could
be distinguishable through the behaviour of the charge gap.

Key questions to address in these compounds concern both the magnetic
ordering and the MIT. In particular, we wish to
understand what sets the direction of the ordered moments in \sriro{},
and \srirob{}.  Beyond this we would like to establish a framework
understanding both the itinerant and localized limits of this family
of compounds. This will provide useful guidance for our discussion of
\emph{engineered phases} constructed from the same basic building
block in Sec. \ref{sec:topological}.

\subsection{Corner shared octahedra with strong spin-orbit coupling}
\label{sec:corner}
The perovskite iridates are built from corner-shared octahedra as
shown for \sriro{}, \srirob{} and orthorhombic \srirop{} in
Fig. \ref{fig:ruddlesden}. The observed MIT
\cite{moon2008dimensionality} as a function of layer count makes it
necessary to discuss both the weak and strong correlation limits. Here,
we present a brief account of the $j=1/2$ spin models at strong
coupling, as well as the tight-binding models at weak coupling that is
applicable to all networks of corner-shared octahedra.  While the real
materials have structural distortions, we begin with the idealized,
undistorted limit.

One simple approach to attacking the model in
Eq. (\ref{eq:full-model}) is to assume the $j=1/2$ and $j=3/2$ bands
to be well separated, neglecting the inter-band interactions. This
projects the multi-orbital problem of Eq. (\ref{eq:full-model}) into a
single-band model of $j=1/2$ electrons with an effective hopping and
Hubbard interaction
\begin{equation}
\sum_{ij} \sum_{\alpha=\pm} t_{ij} \left(
\h{c}_{i\alpha} c^{}_{j\alpha} +
\h{c}_{j\alpha} c^{}_{i\alpha}
  \right) +
  U_{\rm eff} \sum_i n_{i+} n_{i-},
\end{equation}
where $\h{c}_{i\pm}$ creates a $j_z= \pm 1/2$ electron at site
$\vec{r}_i$ and $U_{\rm eff} \sim (U+2U')/3$.  Inversion and
time-reversal symmetry forbid pseudo-spin dependent hoppings. This
type of itinerant model is a useful starting point to describe the
more three-dimensional compounds, such as those based on \srirop{}. Due
to the increased dimensionality, the bandwidth of the $j=1/2$ states
is larger than the quasi-two-dimensional compounds and becomes
comparable to the effective Coulomb $U_{\rm eff}$.

The magnetic ordering in these models is most easily approached from
the strong Mott limit where $U_{\rm eff} \gg t$. Assuming the
nearest-neighbour hoppings are dominant, this yields an
antiferromagnetic Heisenberg model with an effective ${\rm SU}(2)$
pseudo-spin rotation symmetry \cite{jackeli2009mott}
\begin{equation}
\frac{4t^2}{U_{\rm eff}}  \sum_{\avg{ij}} \vec{\pS}_i \cdot \vec{\pS}_j,
\end{equation}
where $\vec{\pS}_i$ is the $j=1/2$ pseudo-spin at site $\vec{r}_i$ and
$t$ is the nearest-neighbour $j=1/2$ hopping.  Going back to
Eq. (\ref{eq:full-model}) and including the $j=3/2$ excited states or
the Hund's coupling spoils this accidental symmetry and generates all
symmetry allowed terms.  For an ideal corner-shared, 180$^\circ$ bond
these include a four-fold rotation about the bond direction, and
several reflections.  The symmetry allowed terms include a Heisenberg
exchange $J$ and a compass-like, pseudo-dipolar coupling $K$
\cite{jackeli2009mott}
\begin{equation}
  \sum_{\avg{ij}} \left[
    J \vec{\pS}_i \cdot \vec{\pS}_j +
    K (\vhat{r}_{ij} \cdot \vec{\pS}_i) (\vhat{r}_{ij} \cdot \vec{\pS}_j),
    \right]
\end{equation}
where $\vec{r}_{ij} \equiv \vec{r}_j - \vec{r}_i$ is the bond
direction.  In concrete models \cite{jackeli2009mott} one finds $K/J
\propto J_H/U$, so we expect the Heisenberg term to dominant, with
only a small pseudo-dipolar coupling $K \ll J$.

The structure of this simple model provides a partial answer to the
experimental questions introduced in Sec. \ref{sec:magnetism}. Namely,
the approximate SU$(2)$ symmetry renders the final pinning of the
magnetic moments sensitive to SU$(2)$ breaking perturbations, such as
the pseudo-dipolar coupling $K$.  Due to this sensitivity, to
understand the fate of these moments we must understand the complete
set of structural distortions present in the perovskite iridates.

\subsection{Distortions and octahedral rotations}
\label{sec:distortions1}
Two relevant types of distortions in the perovskite iridates are
tetragonal distortion and octahedral rotations. The former arises in
the layered compounds such as \sriro{} and \srirob{}, while octahedral
rotations are present in the entire Ruddlesden-Popper series. Due to
the layered structure, the octahedra in \sriro{} and \srirob{} are
elongated along the inter-layer direction, $[001]$
\cite{crawford1994structural,huang1994neutron,subramanian1994single,
cao2002anomalous}.  Such local distortions can be encapsulated in a
residual crystal field potential that split the $j=3/2$ levels into
two Kramers doublets which remain well separated from the higher-lying
doublet for small distortion \cite{khaliullin2005orbital}. This
highest lying doublet plays the role the $j=1/2$ does in the ideal
case and defines our pseudo-spin.
\begin{figure}[tp]
  \begin{subfigure}[b]{0.43\textwidth}
    \begin{overpic}[width=\textwidth]
      {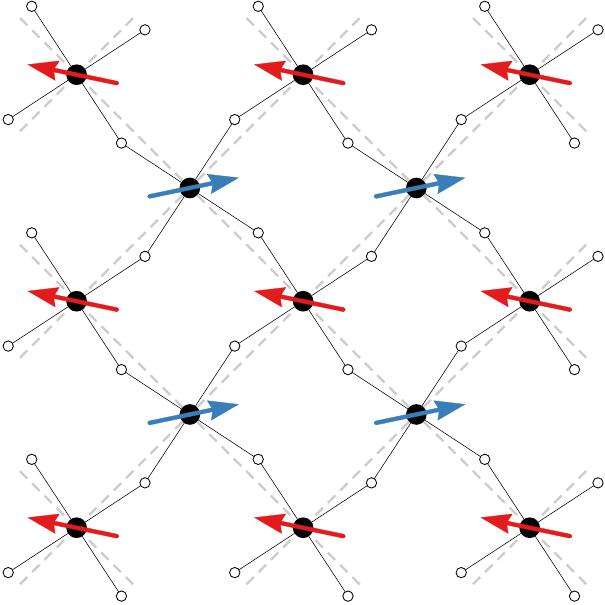}
      \put(22,48){$\theta_{\rm O} \sim 11^{\circ}$}
      \linethickness{0.75pt}
      \put(33,53){\vector(0.5,1.5){4}}
    \end{overpic}
    \caption{\label{fig:rotations} IrO$_2$ layer in \sriro{}}
  \end{subfigure}
  \hspace{1cm}
  \begin{subfigure}[b]{0.15\textwidth}
    \begin{overpic}[width=\textwidth]{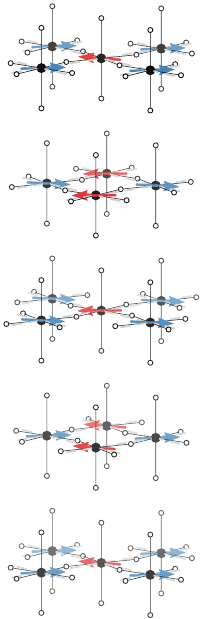}
    \linethickness{0.75pt}
    \put(2,15){\color{black}\vector(0,1){8}}
    \put(2,15){\color{black}\vector(1,0){8}}
    \put(2,15){\color{black}\vector(1,2){2}}
    \put(1.5,23.5){$\scriptsize \vhat{c}$}
    \put(4,19){$\scriptsize \vhat{b}$}
    \put(10.5,14.5){$\scriptsize \vhat{a}$}
    \end{overpic}
    \caption{\label{fig:order-single}\sriro{}}
  \end{subfigure}
  \hspace{0.5cm}
  \begin{subfigure}[b]{0.15\textwidth}
    \includegraphics[width=\textwidth]{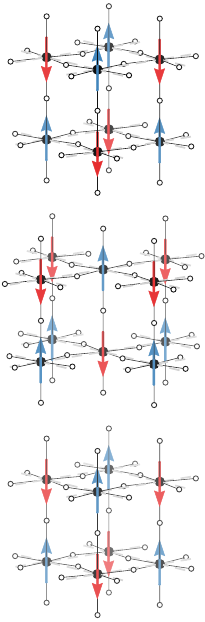}
    \caption{\label{fig:order-bilayer}\srirob{}}
  \end{subfigure}
  \caption{
Illustration of octahedral rotations and magnetic canted
antiferromagnetic order in \sriro{}
\cite{kim2009phase,ye2013magnetic}.  (a) The octahedral rotations and
moment directions in a single IrO\tsub{2} plane of \sriro{}.  (b-c)
Illustration of the stacking of planes and the ordering in (b)
\sriro{} \cite{kim2009phase,ye2013magnetic} and (c) \srirob{}
\cite{fujiyama2012weak}.
  }
\end{figure}
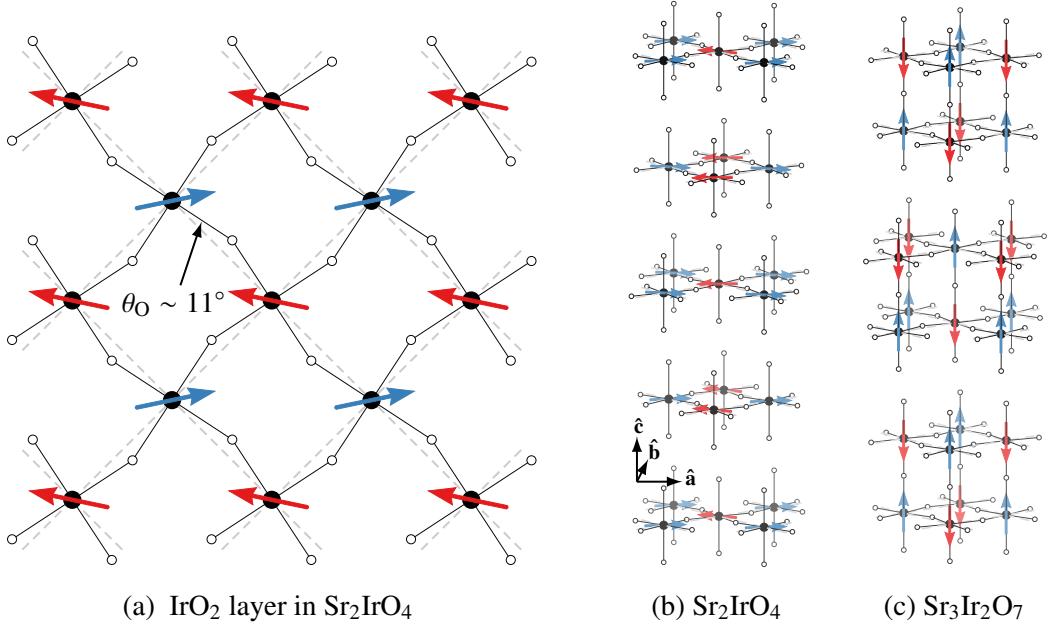
In their simplest form, these distortions manifest as a compression or
elongation along some direction $\vhat{n}$ of the octahedron, modeled
as
\begin{equation}
  \label{eq:distortion}
  V_{\vhat{n}} = {\Delta} \left(\vhat{n} \cdot \vec{L}\right)^2,
\end{equation}
where $\vec{L}$ is the orbital angular momentum projected into the
$t_{2g}$ levels and $\Delta > 0$ corresponds to compression and
$\Delta<0$ to elongation.  For tetragonal distortion $\vhat{n}$ is
along one of the cubic axes, $\vhat{x}$, $\vhat{y}$ or $\vhat{z}$.
Including the SOC, the atomic energy levels can be straightforwardly
found, with the highest lying doublet being a mixture of the $j=1/2$
and $j=3/2$ states as \cite{khaliullin2005orbital}
\begin{eqnarray}
  \ket{\pm} &=& 
\cos{\theta}\ket{\frac{1}{2},\pm \frac{1}{2}} \pm
\sin{\theta}\ket{\frac{3}{2},\pm \frac{1}{2}},
\end{eqnarray}
where the mixing angle is $\tan(2\theta) = 4\sqrt{2} \Delta/(2\Delta +
9\lambda)$.  Including these distortions removes the effective
four-fold rotation symmetry, allowing an additional Ising-like
anisotropy $\sim \Gamma_{zz} \pS^z_i \pS^z_j$ on each bond in the
plane perpendicular to the distortion \cite{jackeli2009mott}.

In addition, octahedral rotations are present and break some of the
rotational and translational symmetry of the ideal lattice, enlarging
the unit cell.  In \sriro{} and \srirob{} the octahedra are rotated
about the $[001]$ axis in a staggered
fashion\cite{crawford1994structural,huang1994neutron,cao2002anomalous},
characterized by a deviation of the bond angle of $ \sim 11^\circ
\equiv \theta_{\rm O}$ from the ideal $180^\circ$, as illustrated in
Fig. \ref{fig:rotations}.  We leave the more involved octahedral
tilting of orthorhombic \srirop{} to the literature
\cite{longo1971structure,zhao2008high}.  Such rotations lower the bond
symmetry, breaking the inversion symmetry about the bond center and
allow pseudo-spin dependent hoppings and Dzyaloshinskii-Moriya (DM)
exchange \cite{jackeli2009mott,wang2011twisted}. Schematically, a
tight-binding model for the $j=1/2$ electrons in a single IrO\tsub{2}
plane that includes such effects is
\begin{equation}
 \sum_{\avg{ij}}\left[ 
-t \sum_{\alpha} \left( \h{c}_{i\alpha} c^{}_{j\alpha} + {\rm h.c.}\right)
+it_z \sum_{\alpha\alpha'}
(-1)^i
\left(\h{c}_{i\alpha} \sigma_z c_{j\alpha'} +{\rm h.c.}\right)
\right],
\end{equation}
where $(-1)^i = \pm 1$ is a staggered sign, as illustrated in
Fig. \ref{fig:rotations} and one expects $t \sim \cos{2\theta_{\rm
O}}$ and $t_z \sim \sin{2\theta_{\rm O}}$.  The pseudo-spin dependent
term $t_z$ can be removed by rotating the pseudo-spin at each site by
an angle $(-1)^i \phi$ about the $\vhat{z}$ axis, where $\tan 2\phi =
t_z/t$ \cite{wang2011twisted}. In these rotated axes the effective
spin model is once again purely Heisenberg \cite{jackeli2009mott},
with $J \sim 4(t^2+t_z^2)/U_{\rm eff}$, without the DM or symmetric
anisotropic exchanges that were expected in terms of the original
pseudo-spins.

This is an accident of the strict $j=1/2$ limit
\cite{jackeli2009mott}; the symmetric anisotropies, such as
$\Gamma_{zz}$ introduced by tetragonal distortion, the pseudo-dipolar
$K$ or the inter-layer anisotropies in the bilayer case will not be
removed.  These break the pseudo-${\rm SU}(2)$ symmetry and thus set a
preferred direction or set of directions in pseudo-spin space.  An
appropriate choice of these terms \cite{jackeli2009mott} can pin the
moment in a direction that is staggered following the octahedral
rotations as is found in \sriro{}.  The case of \srirob{} is less
clear, though the moment orientation being very different is not
surprising in light of the above discussion. However, it is likely
this compound lies far from pseudo-SU$(2)$ invariant limit given the
very large gap seen in the spin-wave spectrum
\cite{kim2012large}. While consensus has not yet been reached, several
proposals have put forth mechanisms to explain this collinear order by
including large anisotropic couplings
\cite{kim2012large,carter2013microscopic}.

\subsection{Topological phases}
\label{sec:topological}
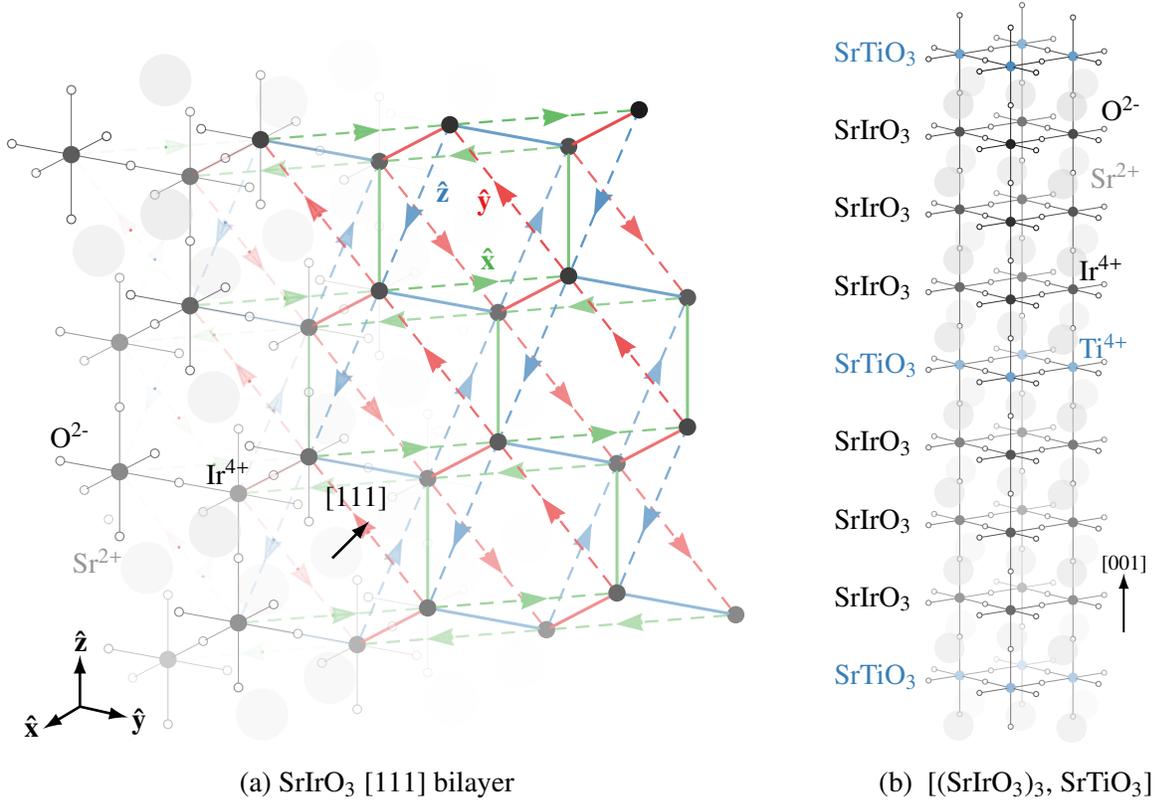
\begin{figure}
  
  \begin{subfigure}[b]{0.6\textwidth}
  \begin{overpic}[width=\textwidth]{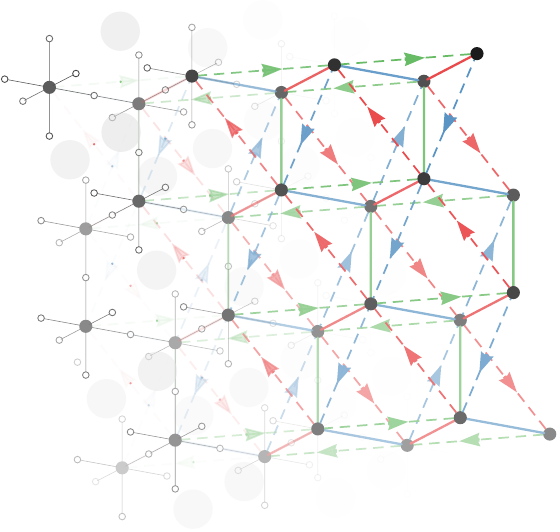}       
    \put(27,35){Ir\tsup{4+}}
    \put(9,23){\textcolor{cgray}{Sr\tsup{2+}}}
    \put(6,40){O\tsup{2-}}
    \put(58,73){$\textcolor{cblue}{\vhat{z}}$}
    \put(63.5,72){$\textcolor{cred}{\vhat{y}}$}
    \put(64,64){$\textcolor{cgreen}{\vhat{x}}$}
    \linethickness{1pt}
    \put(43,32){$[111]$}
    \put(44,25){\vector(1,1){5}}
    \put(10,5){\vector(0,1){7}}
    \put(10,5){\vector(-2,-1.2){5}}
    \put(10,5){\vector(2,-0.45){6.5}}
    \put(9.25,12.5){$\vhat{z}$}
    \put(17,2){$\vhat{y}$}
    \put(2.5,1){$\vhat{x}$}
  \end{overpic}
  \caption{\label{fig:bilayerTI}\srirop{} $[111]$ bilayer}
  \end{subfigure}\hspace{1cm}
  \begin{subfigure}[b]{0.3\textwidth}
    \begin{overpic}[height=2\textwidth]{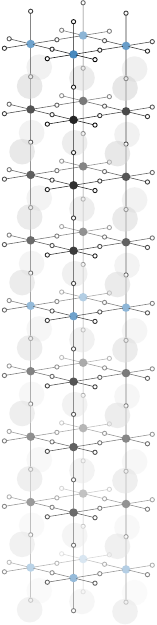}       
      \put(-12,8){\textcolor{cblue}{SrTiO$_3$}}
      \put(-12,18.5){SrIrO$_3$}
      \put(-12,29){SrIrO$_3$}
      \put(-12,39.5){SrIrO$_3$}
      \put(-12,50){\textcolor{cblue}{SrTiO$_3$}}
      \put(21,52){\textcolor{cblue}{Ti\tsup{4+}}}
      \put(21,62.5){Ir\tsup{4+}}
      \put(22.5,75){\textcolor{cgray}{Sr\tsup{2+}}}
      \put(24,84){{O\tsup{2-}}}
      \put(-12,60.5){SrIrO$_3$}
      \put(-12,71){SrIrO$_3$}
      \put(-12,81.5){SrIrO$_3$}
      \put(-12,92){\textcolor{cblue}{SrTiO$_3$}}
      \linethickness{0.75pt}
      \put(27,15){\color{black}\vector(0,1){8}}
      \put(24,23.5){\scriptsize $[001]$}
    \end{overpic}
    \caption{\label{fig:super}
      [(SrIrO$_3$)$_3$, SrTiO$_3$]
    }
  \end{subfigure}
  \caption{
    (a) An ideal bilayer of \srirop{} showing the
    honeycomb lattice formed by the Ir ions. First nearest neighbours
    are shown by solid lines while second nearest-neighbour bonds are
    dashed. The arrow indicates the hopping direction, while the colour
    indicates whether $\vhat{d}_{ij}=\vhat{x}$, $\vhat{y}$ or $\vhat{z}$. 
    (b) An ideal superlattice [(\srirop{})\tsub{3},\srtio{}] along
    the $[001]$ direction. 
  }
\end{figure}

While in the above sections we started from the Mott insulating limit,
in moving to orthorhombic \srirop{} we will start from the itinerant
limit. As bulk \srirop{} is a semi-metal with large SOC, variants of
this compound may be a good place to look for topological insulators
or metals.  A band insulator is deemed a topological insulator (TI),
if it cannot be smoothly deformed into a decoupled, atomic insulator
in such a way that preserves time-reversal
symmetry\cite{qi2011topological,hasan2011three}.  In two or three
dimensions this classification into trivial and non-trivial is
complete without any further subdivisions
\cite{qi2011topological,hasan2011three}.  TIs have a number of very
interesting properties such as gapless edge states and unusual
magneto-electric response \cite{qi2011topological,hasan2011three}.  As
discussed in the introduction, most examples of TIs are found in
weakly correlated systems. Some proposals have been put forward to
look for TIs in transition metal oxides (TMOs): these promise to not
only be more robust, but provide a playground to explore the effects
of interactions on such topological phases
\cite{krempa2014correlated,turner2013beyond}.

Among the iridates, it was suggested that \nairo{} could realize a TI
with the pseudo-spin $j=1/2$ band acquiring a non-trivial Z$_2$
topological index \cite{shitade2009quantum}.  While there were some
early attempts at explaining the magnetism of \nairo{} within a
band-type picture \cite{mazin20122}, experimentally it appears that
\nairo{} is a Mott insulator. In three dimensions, the pyrochlore
iridates were also put forth as candidates to realize a strong
TI\cite{krempa2014correlated}. Another promising approach is to design
the materials of interest by adopting techniques developed in studies
of oxide heterostructures \cite{hwang2012emergent}.  These methods
allow for a wide variety of lattice geometries with a greater control
over structural distortions and impurity content than is available in
bulk samples.  Heterostructures built from $4d$ or $5d$ TMOs, with
their large intrinsic SOC, are thus likely to offer new and
interesting phases for study.

A simple example of such physics can be illustrated via bilayers of
\srirop{} grown along the $[111]$ direction
\cite{xiao2011interface,lado2013ab}.  In these bilayers the Ir atoms
form a buckled honeycomb lattice of corner-shared octahedra, as shown
in Fig \ref{fig:bilayerTI}.  As in the proposals to realize a TI in
\nairo{}, we have an itinerant $j=1/2$ model of the form
\begin{equation} -t \sum_{\avg{ij}} \sum_{\alpha} \left( \h{c}_{i
\alpha} c^{}_{j \alpha} - \hc \right) + t' \sum_{\avg{\avg{ij}}}
\sum_{\alpha\beta} \left[ \h{c}_{i \alpha}
\left(\vec{\sigma}_{\alpha\beta} \cdot \vhat{d}_{ij} \right) c^{}_{j
\beta} + \hc \right] + \epsilon \sum_{i\alpha} (-1)^i \h{c}_{i\alpha}
c^{}_{i\alpha},
\end{equation} 
where $\h{c}_{i\alpha}$ is a creation operator of $j=1/2$ electron
with the pseudo-spin $\alpha \equiv j_z= \pm 1/2$ at site $\vec{r}_i$.
The sums over $\langle ..\rangle$ and $\langle \langle .. \rangle
\rangle$ denote the nearest neighbour (NN) and next nearest neighbour
(NNN) bonds, respectively. On the NNN bonds the absence of inversion
symmetry allows pseudo-spin dependent terms to appear parametrized by
$\vhat{d}_{ij} = \vhat{x}, \vhat{y}$ or $\vhat{z}$ as illustrated in
Fig. \ref{fig:bilayerTI}. Due to the buckling of the honeycomb
lattice, there is an atomic potential $\epsilon$ staggered between the
two sublattices.  With $t'=\epsilon = 0$, this system is a semi-metal
with Dirac cones, as in graphene, while finite $t'$ or $\epsilon$ gaps
out these cones giving a band insulator. When the atomic potential is
sufficiently small, this model supports a TI with a non-trivial $Z_2$
index and the associated gapless edge modes.

The simple example presented above only begins to scratch the surface
of what is possible in these types of engineered systems. One
particular class of system that has been systematically studied is a
series of artificial superlattices [(SrIrO$_3$)$_m$, SrTiO$_3$] where
$m$ is integer, grown along the [001]-axis atop a substrate
\srtio{}. \cite{matsuno2014engineering} These superlattices consist of
atomically thin slices of \srirop{} separated by layers of insulating
\srtio{}.  While bulk \srirop{} forms a post-perovskite structure
under ambient conditions \cite{longo1971structure}, when prepared in
thin films \cite{jang2010electronic} or superlattices
\cite{matsuno2014engineering}, the \srtio{} layers stabilize the
orthorhombic perovskite structure.  By tuning the number of IrO$_2$
layers, an insulator was achieved for $m=1$ single and $m=2$ bilayer
superlattices, and trilayer $m=3$ illustrated in Fig. \ref{fig:super}
sits at the verge of MIT driven by the layer number $m$
\cite{matsuno2014engineering}.  While these $m=1$ and $m=2$
superlattices are topologically trivial $j=1/2$ insulators, a
theoretical study on single and bilayer \srirop{} interleaved with a
band insulator with orthorhombic perovskite structure (such as
GcScO\tsub{3}, CaTiO\tsub{3}, SrZrO\tsub{3} or SrHfO\tsub{3}) suggests
a rich phase diagram including topological magnetic insulators,
topological crystalline insulators and topological valley insulators
\cite{chen2014topological1}.

The stability of some topologically protected states hinges the
presence of symmetries. A richer topological classification has been
recently uncovered by considering the phases protected by the spatial
symmetries of the crystal. These \emph{topological crystalline
insulators} share many of the same features as $Z_2$ TIs, such as
protected gapless surface states \cite{fu2011topological}. In
contrast, in the prototypical Weyl semi-metal,
\cite{wan2011topological} the gapless surface states remain protected
even in the absence of any symmetry. Given the finer classification
that exists for insulators, one can ask if there are topological
\emph{metals} that are protected by spatial symmetries.  A large class
of such topological metals have since been classified
\cite{turner2013beyond}; these go beyond the Weyl semi-metal and have
surface states protected by either global symmetries or crystal
lattice symmetries, or combination of the two. Depending on the
symmetry properties and the dimension of the Fermi surface, these
surface states can form Dirac cones, flat bands or Fermi arc
states.\cite{matsuura2013protected}.  Among the iridates, the three
dimensional perovskite iridates AIrO\tsub{3} with A an alkaline earth
metal have been recently proposed as an example of topological metals
protected by spatial symmetries. These \emph{topological crystalline
metals} are analogous to the topological crystalline insulators, with
crystal symmetry responsible for the protected surface
states. \cite{chen2015topological} While the bulk states show a nodal
of ring of gapless excitations, protected by time-reversal and spatial
symmetries, the associated two-dimensional surface states are flat in
one direction, while linearly dispersing in the other.  If one adds
external symmetry breaking terms such as those that break
time-reversal, mirror, or glide symmetry, this topological metal acts
as a seed for a rich family of topological phases, such as a Weyl
semi-metal, strong or weak topological insulator, or topological
magnetic insulator \cite{chen2015topological}.

\subsection{Superconductivity}
We now return to the original motivation for the study of the iridates
and \sriro{} in particular: possible links to the physics of
high-temperature superconductivity \cite{kim2012magnetic} and the
realization of exotic superconducting pairing symmetry
\cite{khaliullin2004low} in doped spin-orbit Mott insulators.  In
multi-orbital systems it is a challenge to understand the role played
by the orbital degrees of freedom in the microscopic mechanism of
superconductivity.  The superconductor \srruo{} provides a relevant
example, with all three $t_{2g}$ orbitals important in constructing
Fermi surface. Even such a $4d$ transition metal SCs with intermediate
SOC, multi-orbital interactions such as Hund’s coupling can be
important. \cite{georges2013strong} Moving to $5d$ transition metals,
the large SOC can no longer be ignored and the system may display a
complex combination of spin-singlet and spin-triplet SC order
parameters. As the SOC and other electronic interactions such as
Hund's coupling become comparable, the determination of the ground
states in such multi-orbital systems is highly non-trivial. In
particular, we will discuss how the physics of doping in \sriro{} is
both similar to, but different from the case of cuprate
superconductors.  We then discuss some recent theoretical work that
explore possible pairing symmetries and mechanisms for superconductors
arising from doped Mott insulators with strong SOC. From there, we
review the on-going experimental search for superconductivity in doped
iridates, focusing in particular on progress in doping \sriro{}.

\begin{figure}
\begin{subfigure}[b]{0.29\textwidth}
  \includegraphics[width=\textwidth]{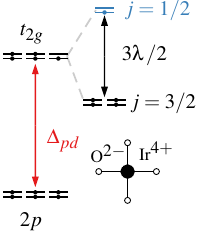}
  \caption{\label{fig:doping} Ir-O energy levels}
\end{subfigure}\hspace{0.5cm}
\begin{subfigure}[b]{0.6\textwidth}
  \includegraphics[width=\textwidth]{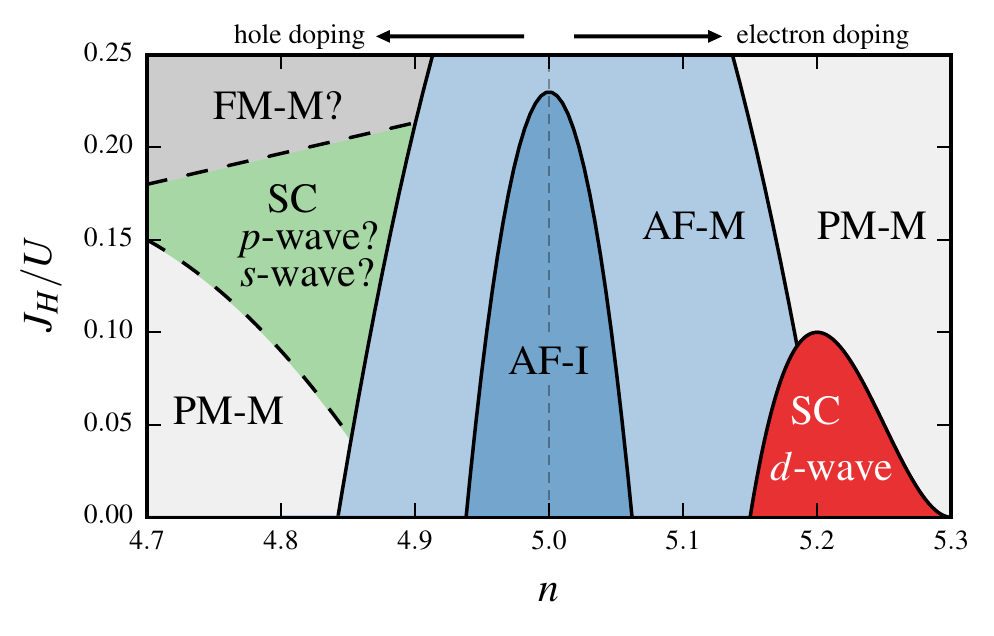}
  \caption{\label{fig:sc} Schematic phase diagram}
\end{subfigure}
\caption{
(a) An illustration of the energy levels for the iridium and oxygen
atoms. A hole on the oxygen is more costly than one in the $j=3/2$
states. (b) Schematic phase diagram for \sriro{} as a function of
filling $n$ and Hund's coupling $J_H$, based on
Ref. [\onlinecite{meng2014odd}]. A variety of phases appear such as an
antiferromagnetic insulator (AF-I) and metal (AF-M), a paramagnetic
metal (PM-M), a ferromagnetic metal (FM-M) as well as possibly two
distinct superconducting (SC) phases.
}
\end{figure}

The physics of doping electrons into these $j=1/2$ Mott insulators is
familiar. Given the filled $2p$ orbitals of the neighbouring oxygens
and the filled $j=3/2$ states, the least costly place to put an extra
electron is in the $j=1/2$ states themselves. The penalty is the
Coulomb energy $U$; much smaller than filling any of the higher-lying
electronic states. We caution that this is purely at the atomic level
and electronic structure effects can alter this identification.  For
example for \sriro{} it has been argued that the electron doping is
similar to the hole-doped cuprates due to an opposite sign in the
next-nearest neighbour hopping integral \cite{wang2011twisted}.
However when iridates are hole-doped, the multi-orbital nature can
become active and it may not be this simple. The key distinction is in
the energy gap to the nearest $5d$ states, the $j=3/2$ levels. In the
cuprates, tetragonal distortion separates the ground state doublet
from the nearest $3d$ levels by a gap of $\sim 1-2\eV$
\cite{lee2006doping}.  This is larger or comparable to the cost of
putting the hole onto the neighbouring oxygens once hybridization
effects are taken into account.  In the case of the iridium oxides the
$j=3/2$ states are only $\sim 0.5\eV$ away while the oxygens are $\sim
1-2\eV$. It is then favourable to put the hole not on the oxygens, but
on \ir{} itself in the $j=3/2$ states. Hund's coupling between the
$j=1/2$ and $j=3/2$ states can further stabilize these holes.  This is
schematically summarized in Fig. \ref{fig:doping}.  A single-orbital
$j=1/2$ picture, enabled by the formation of a Zhang-Rice singlet, may
no longer be tenable and one must move beyond a $t$-$J$ type model and
include aspects of the full multi-orbital problem. This can be seen
clearly in how these same $j=3/2$ excited states appear in the
localized spin model for the half-filled $d^5$ case, where they
generate bond-dependent exchanges as discussed in
Sec. \ref{sec:corner} and will be essential in Sec. \ref{sec:kitaev}.

The classification of pairing terms is also modified when SOC is
strong; as spin is no longer a conserved quantity spin-singlet or
spin-triplet lose their meaning. However a version of this can be
recovered if the strong SOC leads to a single FS with $j=1/2$
character and the multi-orbital problem can be reduced to an effective
single band model.  If the pseudo-spin dependent terms can be made
small, one can define effective pseudo-spin-singlet and
pseudo-spin-triplet SC pairing symmetries.  This assumption is likely
valid in \sriro{}, where most of FS is of $j=1/2$ character and the
nearest-neighbour pseudo-spin dependent terms can be absorbed into the
definition of the $j=1/2$ states \cite{kim2008novel}.  Given the
similarity in lattice structure and Mott physics between \sriro{} and
La$_2$CuO$_4$, it was proposed that a pseudo-spin-singlet $d$-wave
high temperature SC phase like cuprates may emerge in electron-doped
iridates. \cite{wang2011twisted} As discussed above, the hole-doped
case may be significantly different due to the competition between
Hund's coupling and SOC.  With finite Hund's coupling one may expect
an effective interaction through the exchange of ferromagnetic
pseudo-spin fluctuations. So long as the Hund's coupling is not so
large as to drive the system into a ferromagnetic state, one then
would expect to generate a pseudo-spin triplet pairing. Indeed, recent
large-scale dynamical mean-field theory calculations support such a
picture. In these simulations a $d$-wave pseudo-spin-singlet SC was
found on the electron-doped side, while a topological $p+ip$-wave
pseudo-spin-triplet SC was found on the hole doped side
\cite{meng2014odd}. These may be relevant for hole-doped \sriro{}, or
possibly even \srruo{}. A schematic phase diagram as a function of
doping and Hund's coupling is shown in Fig. \ref{fig:sc}.

Experimental studies on physics of doping holes or electrons into
\sriro{} have been carried out through a variety of means; these
include electron doping through depleting oxygen
\cite{korneta2010electron}, La-substitution
\cite{klein2008insight,ge2011lattice}, or surface doping
\cite{kim2014fermi} as well as hole-doping through substitution of Rh
for Ir \cite{klein2008insight,cao2014hallmarks}.  While SC has not
been discovered, in each case modest doping suppresses the magnetic
and strongly affects transport.  Doping the bilayer \srirob{} with
$5\%$ La induces a robust metallic state, with the resistivity showing
a rapid drop below 20$\K$. The magnetic order remains finite despite
strong suppression of $T_c$ \cite{li2013tuning}. This is in contrast
to the case of La-doped \sriro{}, where the magnetic order is
completely suppressed in the metallic state. \cite{ge2011lattice} In
addition to the transport and magnetic properties, angle resolved
photoemission spectroscopy (ARPES) provides interesting insights into
physics of the doped iridates. It was reported that Rh-doped \sriro{}
exhibits a pseudo gap and Fermi arcs \cite{cao2014hallmarks} similar
to the doped cuprates. Similar Fermi arcs were also found in surface
electron doped \sriro{}\cite{kim2014fermi} along with the anomalous
waterfall-like feature in the ARPES spectra \cite{liu2015anomalous}.
Though similar Fermi arc-like features observed in La-doped \srirob{},
have been attributed to Fermi pockets masked by matrix element effects
\cite{he2015fermi}, this does not explain the temperature dependence
of the Fermi arcs in Ref. [\onlinecite{kim2014fermi}], suggesting a
different origin.

\section{Spin liquids and unconventional magnetic orders in honeycomb iridates} 
\label{sec:kitaev}
Despite tremendous efforts, the search for a quantum spin liquid in a
real material remains unresolved. A significant amount of attention in
this search has been directed towards geometrically frustrated
antiferromagnets, such as those on Kagom\'e or triangular
lattices. Many of these systems are described by Heisenberg-like
models, possibly extended with ring exchange terms or with small
anisotropies. While some candidates exist, the combination of
theoretical and experimental uncertainties has made definite
confirmation of the spin liquid state difficult. A promising approach
to circumvent some of these theoretical difficulties is based on
Kitaev's exactly solvable honeycomb model
\cite{kitaev2006anyons}. This highly anisotropic compass model
\cite{nussinov2015compass} is frustrated not by the geometry of the
lattice, but by the intertwining of spatial and spin degrees of
freedom. One can generalize this approach, defining such exactly
solvable models on trivalent lattices in two or three dimensions. The
presence of the solvable point present in the model then provides a
controlled starting point to study the stability of the spin liquid
and possible nearby unusual ordered phases. The challenge is then to
find materials that implement these models.

Here we discuss possible realizations of these Kitaev-type models in
Mott insulators with strong SOC. In two dimensions, there are the
well-studied honeycomb iridates \nairo{} and \aliiro{}, which host
magnetically ordered phases thought to be proximate to the spin liquid
phase; and the honeycomb ruthenium chloride \arucl{}, which has been recently
proposed as a $4d$ analogue. Two three-dimensional
lattices have also been synthesized in polymorphs of the honeycomb
\aliiro{}. These are the hyper-honeycomb \bliiro{} and
stripy-honeycomb \gliiro{}. First, we start with an overview of their
shared basic physics: edge-shared octahedra.

\subsection{Spin-orbit Mott insulators with edge-shared octahedra}
\label{edge-shared}
Finding a system that realizes Kitaev's honeycomb model or its
analogues is difficult; no symmetry principle prevents the
introduction of other interactions, such as Heisenberg exchange
\footnote{Imposing the local plaquette symmetries of the Kitaev model
\emph{does} achieve this, but these are not present in the microscopic
physics.}. The best one can hope for is the \emph{dominance} of Kitaev
exchanges among symmetry allowed interactions.  One route to achieve
this was introduced in the pioneering work of \citet{jackeli2009mott}
based on spin-orbit entangled pseudo-spins in a 90$^\circ$ bonding
geometry. This provides a natural mechanism to generate bond-dependent
Ising interactions \cite{khaliullin2005orbital}, the building block
of the Kitaev model.  For concreteness, we discuss this in the context
of iridium oxides, but translation to related compounds such as
\arucl{} is straightforward.

In contrast to the perovskite iridates discussed in
Sec. \ref{sec:perovskites}, we consider materials where the oxygen
octahedra of neighbouring iridium atoms share an \emph{edge}, rather
than a corner.  In addition, these materials are farther from the
itinerant limit discussed for the perovskite case, lying firmly in the
Mott regime.  We assume that the dominant exchange pathways between
iridium $5d$ orbitals proceed through the oxygen $2p$ orbitals with a
large gap $\Delta_{dp}$ between the $5d$ and $2p$ states.  Integrating
out these oxygen states yields an effective inter-orbital hopping $t
\equiv t^2_{pd\pi}/\Delta_{dp}$ with $t_{dp\pi}$ being the
Slater-Koster $\pi$-overlap and between the $5d$ and $2p$
orbitals. The pair of $5d$ orbitals linked depends on the edge shared;
$d_{yz}$ and $d_{xz}$ orbitals mix on $z$-links, $d_{yz}$ and $d_{xy}$
on $y$-links and $d_{xz}$ and $d_{xy}$ on $x$-links, as shown in
Fig. \ref{fig:honey2D}.  To understand the Mott insulating phase we
will consider the effective model in the strong coupling limit where
$U,J_H,\lambda \gg t$, which can be expressed entirely in terms of the
$j=1/2$ pseudo-spins. The simplest approach is to first project the
multi-orbital model of Eq. (\ref{eq:full-model}) into isolated $j=1/2$
bands prior to taking the strong-coupling limit, as was done in
Sec. \ref{sec:corner}.  One finds that the inter-orbital hoppings $t$
\emph{vanish} when projected into the $j=1/2$ bands, and thus no
exchange is generated. Similarly, even in the full multi-orbital
problem, no exchange arises when Hund's coupling is absent. These
results suggest that in this case the complete multi-orbital nature of
the problem must be considered to account for the interactions between
the $j=1/2$ spins. To include both the $j=3/2$ virtual states and the
effects of Hund's coupling, one must use a more realistic limit where
$U,J_H \gg \lambda$ or $U,\lambda \gg J_H$, obtaining an effective
interaction of Kitaev type
\cite{jackeli2009mott}
\begin{equation} 
\sim -\frac{8 t^2_{} J_H}{3 U^2} \pS^{\gamma}_1 \pS^{\gamma}_2,
\end{equation} where $\gamma$ indicates the type of edge shared by
the neighbouring $j=1/2$ spins $\vec{\pS}_1$ and $\vec{\pS}_2$.

This Hamiltonian is only applicable when the oxygen mediated processes
are dominant and $\Delta_{dp}$ is very large. If other hopping
interactions such as direct $5d$-$5d$ overlap
\cite{chaloupka2010kitaev} or other super-exchange pathways
\cite{chaloupka2010kitaev,chaloupka2013zigzag} are included, then one
expects all symmetry allowed terms to be generated. For a pair of
ideal edge-shared octahedra, the bond symmetry allows two additional
terms: a Heisenberg interaction $J$ and a symmetric off-diagonal
exchange $\Gamma$ \cite{rau2014generic}. An effective exchange model
for the pair of spins is then
\begin{equation}
  \label{hkg}
    J \vec{\pS}_1 \cdot \vec{\pS}_2 +
    K \pS^{\gamma}_1 \pS^{\gamma}_2 +
    \Gamma \left(
      \pS^{\alpha}_1 \pS^{\beta}_2 + \pS^{\beta}_1 \pS^{\alpha}_2
    \right),
\end{equation}
where $\alpha$, $\beta$ indicate the two spin directions not equal to
the edge-type $\gamma$. If we assume the exchange physics is local to
the pair of octahedra, then fixing the parameters on one edge is
sufficient to determine the others related by symmetry.  Details of
the dependence of $J$, $K$ and $\Gamma$ on the atomic interactions for
two of the limiting schemes described here can be found in
\citealt{rau2014generic}.

\subsection{Distortions}
The crystal structure of the honeycomb compounds \nairo{} and
\aliiro{}, and the three-dimensional analogues \bliiro{} and \gliiro{}
deviate from this idealized picture due to the presence of trigonal
and monoclinic distortions.  Following the discussion in Section
\ref{sec:distortions1}, for trigonal distortion, in
Eq. (\ref{eq:distortion}) one has $\vhat{n} \equiv
(\vhat{x}+\vhat{y}+\vhat{z})/\sqrt{3}$ or one of its equivalents.  In
the absence of SOC, this splits the $t_{2g}$ levels into an $a_{1g}$
singlet and $e_g$ doublet. When SOC is present, the highest-lying
Kramers doublet connects to the $j=1/2$ states and can be found
explicitly in Ref. [\onlinecite{khaliullin2005orbital}].  Monoclinic
distortion cannot lift the degeneracy any further, but will complicate
the structure and analysis of these ground state wave-functions.
Corrections to interactions between the $j=1/2$ spins will be
introduced through these changes in the atomic wave-functions, as well
as through the kinetic parts of the exchange processes which depend on
the geometry of the oxygen ions.

Rather than attempt to connect the details of the distortions to the
exchange interactions directly, we consider the new interactions
allowed by these reductions of symmetry. In the materials under
consideration, there are two or more symmetry inequivalent sets of
nearest-neighbour bonds, one with higher symmetry and the rest with
lower symmetry.  The higher bond symmetry groups are $222$ for
\bliiro{} and \gliiro{} and $2/m$ for \aliiro{} and \nairo{}. For a
bond with $2/m$ symmetry, the allowed exchanges include those in
Eq. (\ref{hkg}) but with an additional symmetric off-diagonal exchange
\begin{equation}
  \label{gammap}
  \Gamma' \left(
\pS^{\alpha}_1 \pS^{\gamma}_2 + \pS^{\beta}_1 \pS^{\gamma}_2+
\pS^{\gamma}_1 \pS^{\alpha}_2 + \pS^{\gamma}_1 \pS^{\beta}_2
  \right).
\end{equation}
As an example, trigonal distortion of the oxygen octahedra can
introduce such a term \cite{rau2014trigonal}.  For a bond with $222$
symmetry, such a term is forbidden, but due to the lack of inversion
about the bond center an analogous DM
interaction is allowed
\begin{equation}
  D \left(
\pS^{\alpha}_1 \pS^{\gamma}_2 + \pS^{\beta}_1 \pS^{\gamma}_2-
\pS^{\gamma}_1 \pS^{\alpha}_2 -\pS^{\gamma}_1 \pS^{\beta}_2
  \right).
\end{equation}
The lower
symmetry bonds of \aliiro{}, \bliiro{}, and \nairo{} have only inversion
about the bond center while that of \gliiro{} has no symmetry at all. In the
compounds with an inversion, this implies the absence of DM
interaction on these bonds.

Further neighbour exchange interactions can be analyzed in a similar
fashion, though the number of allowed interactions grows
quickly. These are thought to be important
\cite{kimchi2011kitaev,katukuri2014kitaev,yamaji2014first,sizyuk2014importance}
in \nairo{} where the Mott gap is not too large \cite{comin20122}.
Some discussion of the symmetry allowed interactions in \nairo{} for
second and third nearest neighbours can be found in
\citet{yamaji2014first} and \citet{sizyuk2014importance}. While there
has been some work on further neighbour interactions in the
three-dimensional compounds \cite{lee2014emergent,kimchi2014unified},
their effects and importance remain largely unaddressed.

\subsection{Kitaev physics in two dimensions}
\label{kitaev-two-dimensions}
With this framework in hand, we first turn to possible realizations of
Kitaev's two-dimensional honeycomb model. Promising candidates thought
to be proximate to this physics are the honeycomb iridates \nairo{}
and \aliiro{}, and more recently, \arucl{}. These materials do
\emph{not} have a spin liquid ground state, but instead order
magnetically as temperature is lowered. To understand if the proximity
to Kitaev physics is governing their behaviour we first study
these magnetic phases.

\subsubsection{Experimental review}
\label{sec:exp2d}
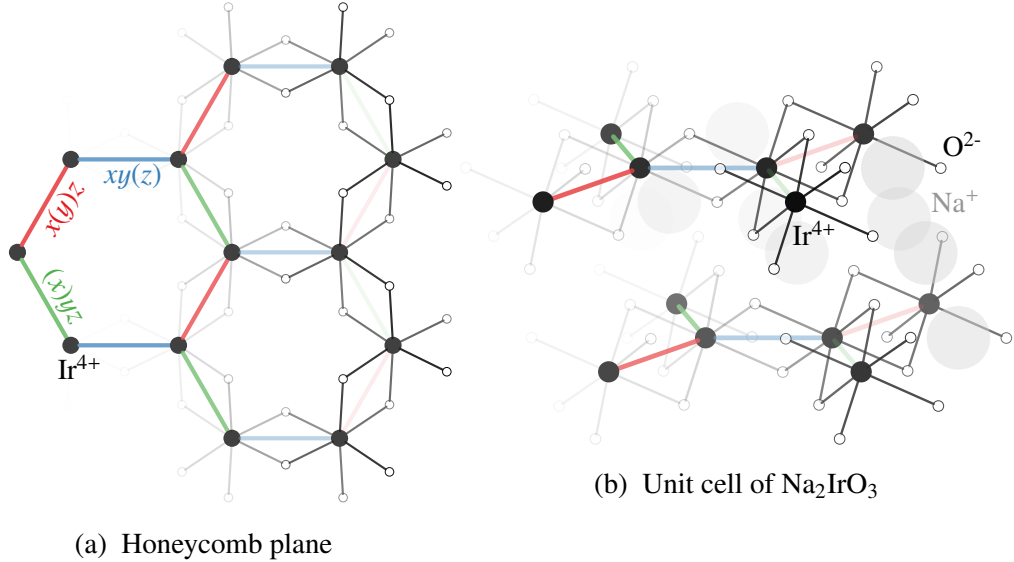
\begin{figure}[tp]
  \begin{subfigure}{0.4\textwidth}
    \begin{overpic}[width=\textwidth]{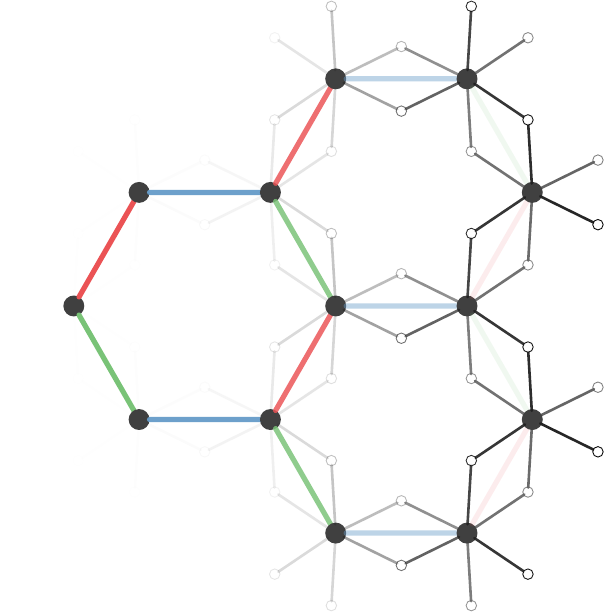}
     \put(29,64){$\textcolor{cblue}{xy(z)}$}
      \put(16,53){\rotatebox{60}{$\textcolor{cred}{x(y)z}$}}
      \put(16,46){\rotatebox{-60}{$\textcolor{cgreen}{(x)yz}$}}
      \put(20,25){Ir\tsup{4+}}
    \end{overpic}
    \caption{\label{fig:honey} Honeycomb plane}
  \end{subfigure} 
  \begin{subfigure}{0.45\textwidth}
    \begin{overpic}[width=\textwidth]{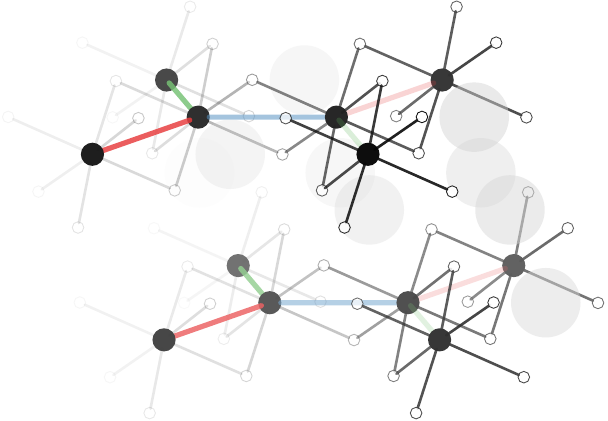}
      \put(60,36){Ir\tsup{4+}}
      \put(85,42){\textcolor{cgray}{Na\tsup{+}}}
      \put(87,52){O\tsup{2-}}    
    \end{overpic}
    \caption{\label{fig:honeyuc} Unit cell of \nairo{}}
  \end{subfigure} 
  \caption{
    \label{fig:honey2D}
Crystal structure of the layered \nairo{}. (a) The
unit cell of \nairo{} showing the quasi-two dimensional layered
structure. (b) The honeycomb plane of \nairo{} showing the labeling
of the nearest neighbour bonds, with a $\gamma=x,y,z$ type bond labeled as
$\alpha\beta(\gamma)$. 
}
\end{figure}

The crystal structures of these compounds (see Fig. \ref{fig:honey2D})
takes the form of layers of edge-shared IrO$_6$ octahedra arranged in
a honeycomb lattice
\cite{singh2010antiferromagnetic,singh2012relevance}.  These octahedra
are compressed in the direction perpendicular to the honeycomb planes,
with further monoclinic distortions along a preferred axis
\cite{ye2012direct,choi2012spin}.  A $340 \meV$ optical gap
\cite{comin20122} and estimates of $3\lambda/2 \sim 0.5 \eV$
\cite{gretarsson2013crystal} for SOC identify \nairo{} as a candidate for a
$j=1/2$ spin-orbit Mott insulator.  Fits of the magnetic
susceptibility confirm the effective spin-1/2 picture, giving a
magnetic moment $\sim 1.82 \mu_B$ with large antiferromagnetic
Curie-Weiss temperature $\theta_{\rm CW} \sim -116 \K$
\cite{singh2010antiferromagnetic}. The low temperature
antiferromagnetic ordering transition seen near $T_{\rm N} \sim 15 \K$
suggests substantial frustration. Neutron scattering and resonant
X-ray scattering studies \cite{liu2011long,ye2012direct,choi2012spin}
have unambiguously identified this ordering as having a \emph{zigzag}
structure.  This zigzag order consists of spins aligned
ferromagnetically along one direction, with the chains alternating
antiferromagnetically as illustrated in Fig.
\ref{fig:hk-phase-diagram}. Dynamical probes, such as inelastic
neutron scattering (INS) and resonant inelastic X-ray scattering
(RIXS), have provided additional clues via access to their
excitations. RIXS studies have shown \cite{gretarsson2013magnetic}
that there is a branch of magnetic excitations at high-energy near
$\sim 30\meV$. Though some dispersion can be identified, any
information about the excitations below $\sim 10\meV$ is lost due to
limitations in energy resolution. INS studies offer a complementary
picture, but due to the large neutron absorption cross-section of Ir,
they have been limited to powder samples \cite{choi2012spin} providing
a higher energy resolution but only a few details of the distribution
in wave-vector.  Nonetheless, below the N\'eel temperature, two key
features can be resolved: scattering near the magnitude of the zigzag
ordering wave-vector is present down to at least $2\meV$ and there is
an absence of scattering at small wave-vectors and energy.  Recent
diffuse magnetic X-ray scattering \cite{chun2015direct} in the
paramagnetic phase has provided an experimental confirmation of
dominant Kitaev interactions, validating the theoretical arguments of
Sec. \ref{edge-shared}. These measurements show a clear locking of the
spatial and spin orientations characteristic of the bond-dependent
Kitaev exchange.  For \aliiro{} the situation is less
clear. Experimentally, an antiferromagnetic ordering transition
\cite{singh2012relevance} is seen at $\sim 15\K$, as in \anairo, but
\aliiro{} has a smaller Curie-Weiss temperature of $\theta_{\rm CW}
\sim -33\K$.  While there have been reports of an incommensurate
ordering wave vector lying in the first Brillouin
zone \footnote{
R. Coldea, Talk at the SPORE13 conference held at MPIPKS,
Dresden (2013); S. Choi, Talk at the APS March
meeting, Denver, CO (2014)
},
the details of the magnetic ordering pattern and the structure of the low
energy excitations remain largely unresolved.

A number of studies have tried to elucidate the properties of these
materials more indirectly through elemental substitution.  One
promising approach is to dope isoelectronically from \nairo{} to
\aliiro{} as \naliiro{}.  For $x \lesssim 0.25$, uniform solid
solutions can be obtained, with both $T_N$ and $|\theta_{\rm CW}|$
suppressed with increased doping
\cite{cao2013evolution,manni2014isoelectronic}. While a quantum
critical point near $x_c \sim 0.75$ was found in
Ref. [\onlinecite{cao2013evolution}], indications of phase separation
reported for the range $0.25 \lesssim x \lesssim 0.6$
\cite{manni2014isoelectronic} complicate this identification.  Further
study is needed for $x \gtrsim 0.6$ near the \aliiro{} end of this
range to clarify the issue.  Another approach is dilution of \ir{}
with non-magnetic Ti$^{4+}$ \cite{manni2014nonmagnetic}, forming
\nairtio{} or \liirtio{}. As the magnetic lattice of Ir is depleted,
both systems enter a spin-glass phase, consistent with the high
frustration indicated by $T_N/|\theta_{\rm CW}|$. In both compounds
the spin glass ordering temperature $T_g$ decreases roughly linearly
as a function of $x$ until to the site-percolation threshold at $x_p
\sim 0.3$. These two cases are
distinguished by the dependence of the Curie-Weiss temperature on
dilution: the \nairtio{} shows a marked decrease as $x$ is increased
while for the \liirtio{} it is essentially constant. This has been
interpreted \cite{manni2014nonmagnetic}
as evidence of more significant long range interactions in \aliiro{}
compared to \nairo{}.

\subsubsection{Theory}
The balance of experimental evidence has suggested that large Kitaev
interactions are necessary to understand the physics of \nairo{} and
\aliiro{}. However, the low-energy physics and the ground state
selection are dependent on the details of the perturbations that take
us away from the Kitaev limit.  These questions have led to a number
of theoretical proposals that have been put forth to explain the
appearance of zigzag ordering in \nairo{} and the nature of the
magnetic ordering in \aliiro{}.
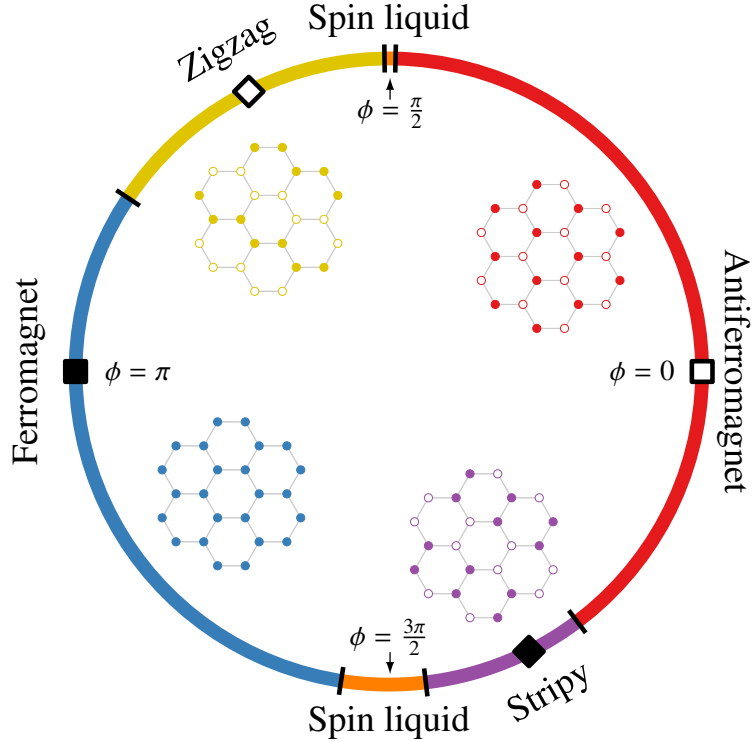
\begin{figure}[t]
  \centering
  \begin{overpic}[width=0.6\textwidth]{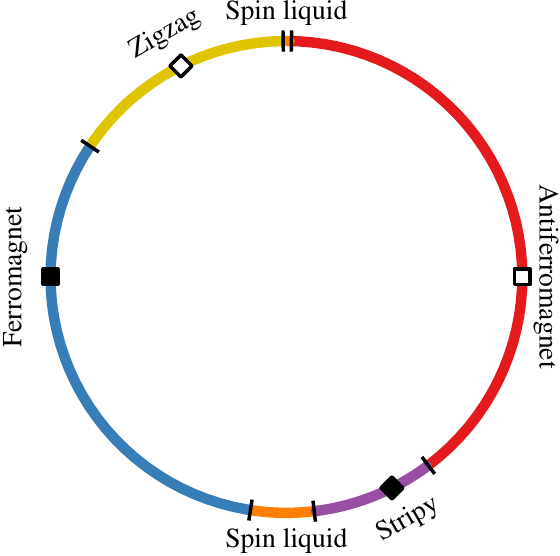}
    \put(20,23){\includegraphics[width=0.12\textwidth]{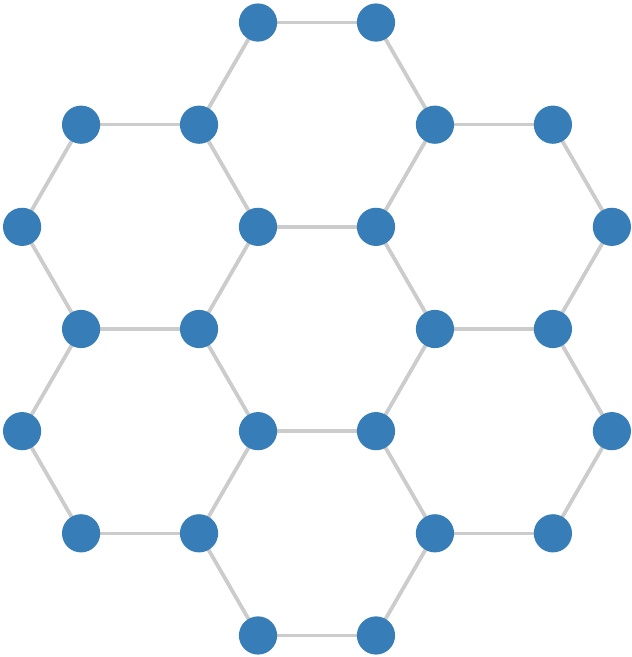}}
    \put(63,55){\includegraphics[width=0.12\textwidth]{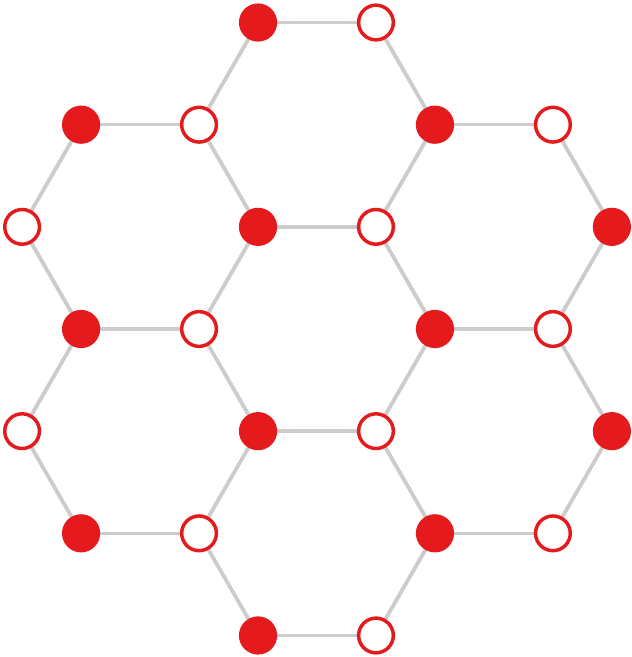}}
    \put(54,16){\includegraphics[width=0.12\textwidth]{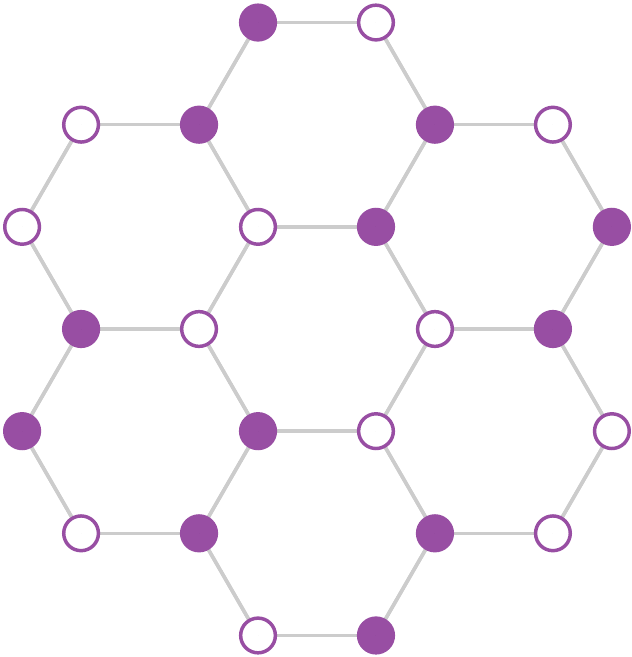}}
    \put(25,60){\includegraphics[width=0.12\textwidth]{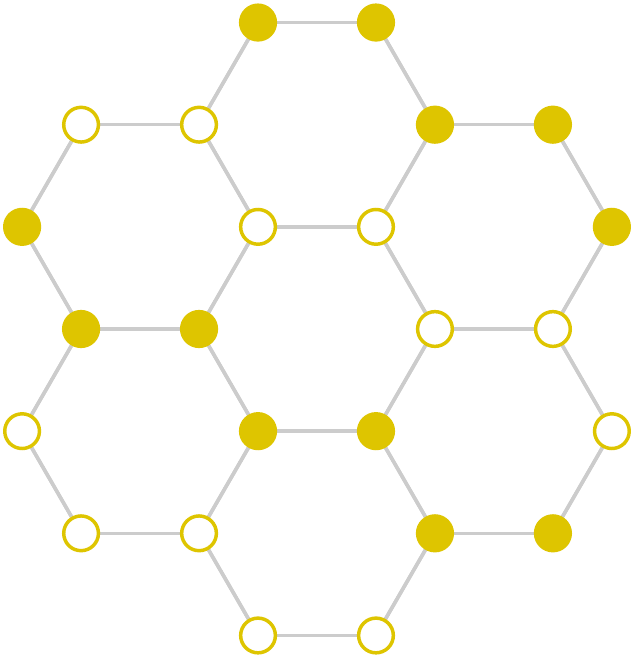}}
    \put(81,48.75){$\phi = 0$}
    \put(47,84){$\phi = \tfrac{\pi}{2}$}
    \put(13,48.75){$\phi = \pi$}
    \put(46,13){$\phi = \tfrac{3\pi}{2}$}
    \linethickness{0.5pt}
    \put(51.4,12){\vector(0,-1){3}}
    \put(51.4,86.5){\vector(0,1){3}}
  \end{overpic}
 \caption{
\label{fig:hk-phase-diagram}
Phase diagram of the Heisenberg-Kitaev model. The model
is parametrized as $J=\cos\phi$, $K=\sin\phi$, as discussed in the main text. Squares
and diamonds show points related by the Klein duality transformation. 
Phases of the Heisenberg-Kitaev model are shown inside, these are
 the ferromagnet (FM), 
antiferromagnet (AFM), stripy (ST) and zigzag (ZZ).
}
\end{figure}

Early work on \nairo{} \cite{chaloupka2010kitaev} primarily focused on
the Heisenberg-Kitaev (HK) model and its extensions. In the HK model,
the Kitaev exchange is supplemented by a conventional Heisenberg
interaction
\begin{equation}
  \sum_{\avg{ij} \in \gamma} \left(J \vec{\pS}_i \cdot \vec{\pS}_j + K S^{\gamma}_i  S^{\gamma}_j
    \right).
\end{equation}
This model has since attracted much theoretical attention, not only
due to its relation to the honeycomb iridates, but also as a model
system to study the stability of the Kitaev spin liquid and its
neighboring phases
\cite{jiang2011possible,reuther2011finite,price2012critical,
you2012doping,schaffer2012quantum}.  The remainder of the phase
diagram of this model can be understood almost completely with the
help of a four-sublattice spin rotation---the so-called \emph{Klein
duality}. This duality maps the HK model to itself, but with the
modified parameters $J' = -J$ and $K' = K+2J$.  Applying this duality
to the ferromagnetic and antiferromagnetic Heisenberg limits yields
two new well-understood limits. These are the stripy (ST) phase at $K
= -2J < 0$ dual to the ferromagnet (FM) and the zigzag (ZZ) phase at
$K = 2J > 0$ dual to the antiferromagnet (AFM). These ordered phases
are illustrated in Fig \ref{fig:hk-phase-diagram}.  The Kitaev spin
liquid is stable to finite $J$ and maps to itself under the Klein
duality.  These five phases encompass all phases present in the
model. This has been borne out using a number of theoretical methods
\cite{chaloupka2010kitaev,jiang2011possible,reuther2011finite}; the
full phase diagram parametrized as $J = \cos{\phi}$, $K=\sin{\phi}$ is
presented in Fig. \ref{fig:hk-phase-diagram}.

For the presumed ferromagnetic Kitaev interaction ($K<0$), the
neighboring magnetic phases are a ferromagnet and a stripy phase
\cite{chaloupka2010kitaev}.  It is possible to obtain a zigzag ground
state with this model, but only when the Kitaev interaction is
antiferromagnetic ($K>0$) and the Heisenberg interaction is
ferromagnetic ($J<0$). The need for a large antiferromagnetic Kitaev
interaction is at odds with expectations from a picture of dominant
oxygen-mediated super-exchange. Some efforts have been made to justify
\cite{chaloupka2013zigzag} this parameter regime, while others
\cite{kimchi2011kitaev,choi2012spin,singh2012relevance} have added
additional couplings to the ferromagnetic Kitaev limit in an attempt
to resolve the discrepancy.  Recent \emph{ab-initio} calculations
\cite{katukuri2014kitaev,yamaji2014first} have supported the view that
ferromagnetic Kitaev interactions are the proper starting point.
Starting from this Kitaev limit, theoretical studies have considered
the effects of the symmetry allowed nearest-neighbour interactions
\cite{rau2014generic,katukuri2014kitaev,rau2014trigonal}, such as the
$\Gamma$ from Eq. (\ref{hkg}) and $\Gamma'$ from Eq. (\ref{gammap}),
as well as further neighbour interactions
\cite{yamaji2014first,sizyuk2014importance}.  While the details
differ, in all of the above works a zigzag phase can be stabilized
near the ferromagnetic Kitaev limit by some combination of further
neighbour and anisotropic interactions.

Even with the scarcity of experimental information, several theories
have been put forth to explain the possibility of incommensurate
ordering in \aliiro{}. A common approach is to move away from some of
the limiting regimes studied for \nairo{}. For example, in
Ref. [\onlinecite{reuther2011finite}], the HK model was extended to
include second nearest-neighbour Heisenberg and Kitaev exchange
interactions. In this regime one can find an incommensurate spiral
phase with wave-vector lying in the first Brillouin zone, as has been
reported in \aliiro{}. In Refs. [\onlinecite{rau2014trigonal}] and
[\onlinecite{chaloupka2015hidden}], one perturbs away from the zigzag
phase stabilized by anisotropic exchanges $\Gamma$ and $\Gamma'$, 
which is relevant for
\nairo{}, to find similar incommensurate phases.  Other approaches
begin from a more monoclinic limit, considering quasi-one-dimensional
chains \cite{kimchi2014unified} or strengthening Kitaev interactions
on one set of bonds \cite{nishimoto2014strongly}. The lack of details
on the reported incommensurate spiral phase coupled with the same
large parameter space that plagues \nairo{} leaves many questions open
for \aliiro{}.

\subsubsection{\arucl{}}
The $4d$ compound \arucl{} has recently attracted attention as another
possible system to explore Kitaev physics. As in honeycomb iridates,
the Ru$^{3+}$ ion has a $d^5$ configuration but with an octahedral
cage of Cl$^{-}$ rather than O$^{2-}$. These octahedra are then
arranged in a layered, edge-shared honeycomb network with space group
$P3_112$ \cite{stroganov1957}.  While first identified as a band
insulator \cite{binotto1971optical}, spectroscopic studies
\cite{pollini1996electronic, plumb2014alpha} have since converged to a
picture of \arucl{} as a Mott insulator. Indeed, this is supported by
measurements of the magnetic susceptibility, which takes a Curie-Weiss
form with an average moment $\sim 2.0 -2.3\ \mu_B$, not too far from
what would be expected for a spin-1/2 \cite{fletcher1967x,
kobayashi1992moessbauer}. Going to lower temperatures,
antiferromagnetic order appears near $\sim 7\K$, as seen in specific
heat and magnetic susceptibility measurements
\cite{kubota2015successive, sears2015magnetic,
majumder2015anisotropic}. Neutron scattering experiments
\cite{sears2015magnetic, banerjee2015proximate} have identified this
order as being zigzag \cite{sears2015magnetic, banerjee2015proximate}
as seen in \anairo{}.  Further, a broad specific heat feature seen
near $\sim15\K$ \cite{kubota2015successive, sears2015magnetic} may be
an indication of a multi-step transition process.

While strongly suggestive of the same physics seen in the honeycomb
iridates, one may be concerned with the applicability of the $j=1/2$
picture itself.  Given the considerably smaller SOC in the $4d$
orbitals of Ru$^{3+}$ \cite{sandilands2015orbital}, strong overlap of
the $j=1/2$ and $j=3/2$ bands is possible and would require a more
complicated description of the Mott insulating phase. One proposed
resolution \cite{shankar2014kitaev}, motivated by ab-initio
calculations, invokes the enhancement of SOC by electronic correlation
effects. This SOC enhancement cleanly separates the $j=1/2$ states from
the $j=3/2$ and increases the tendency towards a Mott phase. Several
theoretical descriptions \cite{shankar2014kitaev,
kubota2015successive, banerjee2015proximate} have been put forth to
explain the appropriate low-energy physics, as well as the nature of
the ordered phase that appears at low temperature.

Some of the key differences between \arucl{} and the honeycomb
iridates may render it a much better system to study Kitaev
magnetism. Indeed, geometrically the local environment of the Ru ion
is nearly distortion-free \cite{plumb2014alpha,sears2015magnetic},
lacking the significant trigonal or monoclinic distortions that have
been reported \cite{singh2010antiferromagnetic,singh2012relevance} in
\anairo{} and \aliiro{}.  Access to the details of the magnetic
excitation spectrum, due to the absence of neutron absorbing Ir atoms
\cite{banerjee2015proximate}, should further illuminate the physics of
this material. Aside from pinning down the theoretical description,
with more information and fewer parameters it may be easier to perturb
the system toward the sought-after Kitaev spin liquid phase
\cite{banerjee2015proximate}.

\subsection{Kitaev physics in three dimensions}
\label{sec:kitaev-three-dimensions}
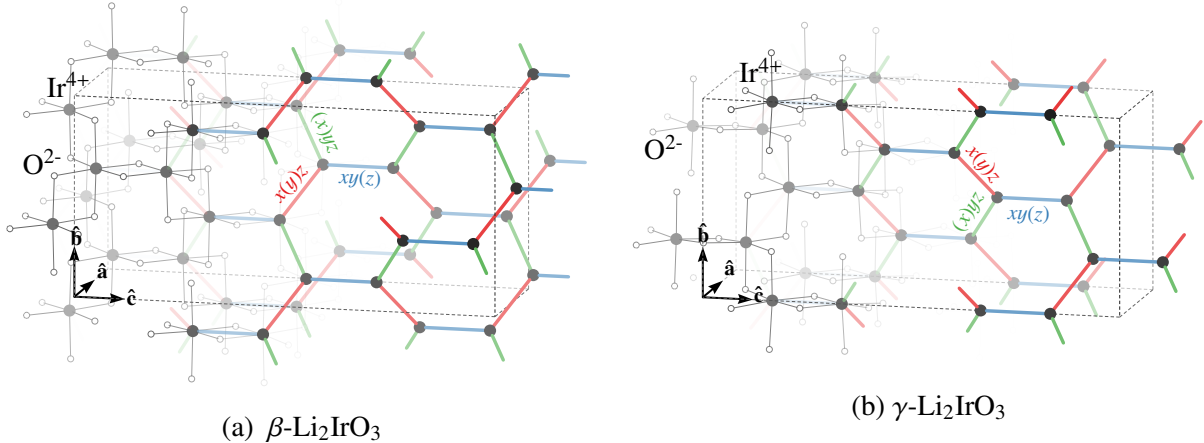
\begin{figure}[tp]
  \begin{subfigure}{\textwidth}
    \centering
    \begin{subfigure}{0.48\textwidth}
      \begin{overpic}[width=\textwidth]{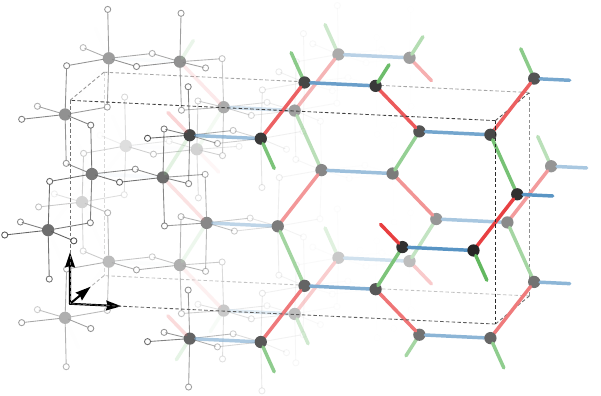}
        \put(20.5,14){$\scriptsize \vhat{c}$}
        \put(15.5,18.5){$\scriptsize \vhat{a}$}
        \put(11,24){$\scriptsize \vhat{b}$}
        \put(7,49){Ir\tsup{4+}}
        \put(3,36){O\tsup{2-}}
        \put(56,34.5){\rotatebox{-3}{\scriptsize $\textcolor{cblue}{xy(z)}$}}
        \put(44.5,31){\rotatebox{50}{\scriptsize $\textcolor{cred}{x(y)z}$}}
        \put(51,46){\rotatebox{-65}{\scriptsize $\textcolor{cgreen}{(x)yz}$}}
      \end{overpic} 
      \caption{\label{fig:hyperhoney} \bliiro{}}
    \end{subfigure}  \hspace{0.25cm}
    \begin{subfigure}{0.48\textwidth}
      \begin{overpic}[width=\textwidth]{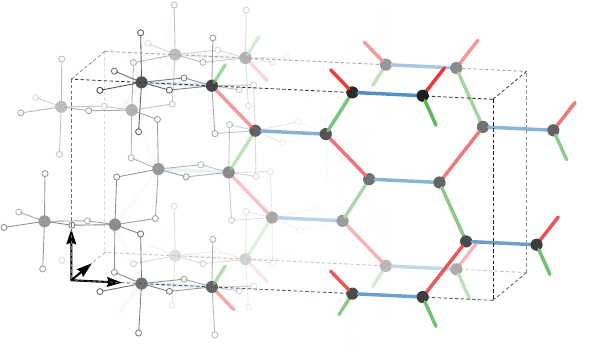}
        \put(20.5,11){$\scriptsize \vhat{c}$}
        \put(15.5,15.5){$\scriptsize \vhat{a}$}
        \put(11,21){$\scriptsize \vhat{b}$}
        \put(18,48){Ir\tsup{4+}}
        \put(2,35){O\tsup{2-}}
        \put(63,25){\rotatebox{-3}{\scriptsize $\textcolor{cblue}{xy(z)}$}}
        \put(55.5,36.5){\rotatebox{-50}{\scriptsize $\textcolor{cred}{x(y)z}$}}
        \put(53.5,22.5){\rotatebox{60}{\scriptsize $\textcolor{cgreen}{(x)yz}$}}
      \end{overpic}
      \caption{\label{fig:harmonichoney}\gliiro{}}
    \end{subfigure} 
  \end{subfigure}
  \caption{
Crystal structures of the (a) hyper-honeycomb \bliiro{} and the (b)
stripy-honeycomb \gliiro{}.  Each structure is formed from a network
of approximately edge-shared IrO$_6$ octahedra, with bonds of type
$\gamma=x,y,z$ labeled as $\alpha\beta(\gamma)$.
\label{fig:hypers}
}
\end{figure}

The search for spin liquids in three dimensional materials has focused
on a handful of frustrated lattices, for example on the pyrochlore
\cite{canals1998pyrochlore, gingras2014quantum} or the hyper-kagome,
realized in the iridate \hknairo{}
\cite{okamoto2007spin,singh2013spin}. As in the two-dimensional cases
discussed in Section \ref{sec:kitaev} have been put forth, the
definite confirmation of the spin liquid state remains very
challenging both theoretically and experimentally. While further work
is warranted, following our discussion in two-dimensions, an
alternative route to a three-dimensional spin liquid might lie through
an extension of the Kitaev's honeycomb model.

The exact solvability of Kitaev's 2D honeycomb model depends on the
three-fold coordination of each site and the appropriate assignment of
Ising components among the bonds, \textit{i.e.} the Ising interactions
are along the $x$, $y$, or $z$ directions while bonds with the same
Ising components do not share the same site.  These properties can be
realized on many lattices, hence Kitaev's 2D honeycomb model can be
generalized to three-dimensions. One such example is a spin-$1/2$
model defined on a 3D deleted-cubic lattice, whose exact spin liquid
solution and excitations were studied in
Ref. [\onlinecite{mandal2009exactly}].  Fortuitously, an iridate material with
the Ir ions residing on a topologically-equivalent lattice was
synthesized several years later: the hyper-honeycomb
\bliiro{}.\cite{takayama2015hyperhoneycomb} Almost simultaneously, another
polymorph---the stripy-honeycomb \gliiro{}---was discovered
independently.\cite{modic2014realization} Both polymorphs contain the
correct trivalent lattice structure, bond geometry, and large SOC to
possibly realize a Kitaev model, thereby providing a new avenue toward
the discovery of a 3D spin liquid.  However, magnetic orders were
experimentally observed in both \bliiro{} and
\gliiro{}.\cite{biffin2014unconventional, biffin2014noncoplanar}
Similar to the two-dimensional case, both theoretical and experimental
work has focused on the magnetic phases in order to understand the
role played by Kitaev physics.  Additionally, theoretical
investigations have been undertaken to explore this new 3D spin liquid
appearing in the Kitaev limit.

\subsubsection{Experimental review}

\begin{figure}[tp]
  \begin{subfigure}{\textwidth}
    \centering
    \begin{subfigure}{0.6\textwidth}
      \begin{overpic}[height=0.5\textwidth]{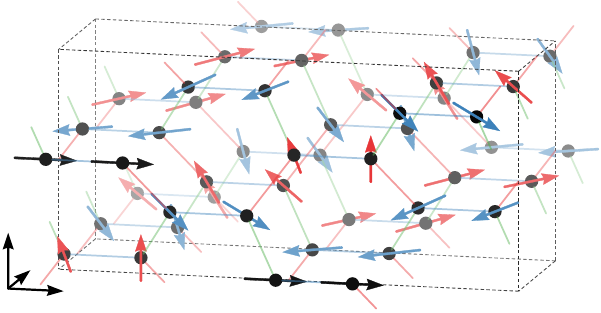}
        \put(11,3){$\scriptsize \vhat{c}$}
        \put(5,7.5){$\scriptsize \vhat{a}$}
        \put(0.5,14){$\scriptsize \vhat{b}$}
        \put(5,50){$\textcolor{clgray}{\vhat{Q}}$}
        \linethickness{3pt}
        \color{clgray}
        \put(5,45){\vector(1.2,1){10}}
      \end{overpic}
    \end{subfigure}  \hspace{0.25cm}
    \begin{subfigure}{0.3\textwidth}
      \begin{overpic}[width=\textwidth]{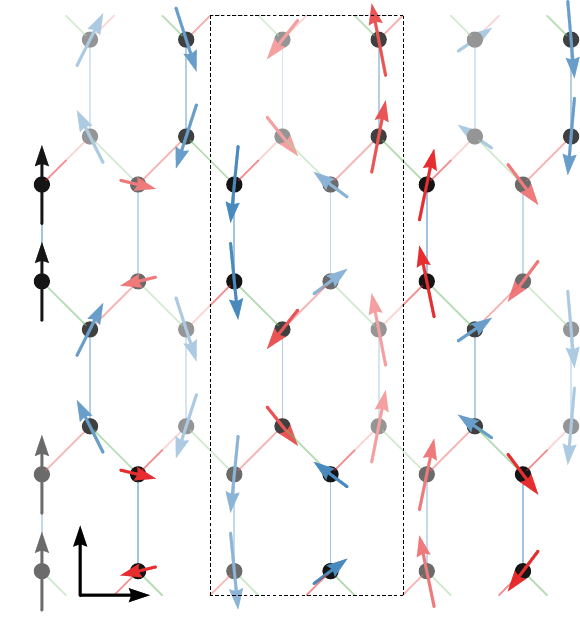}
        \put(24.5,4.5){$\scriptsize \vhat{a}$}
        \put(11.5,18){$\scriptsize \vhat{c}$}
        \put(68,-7){$\textcolor{clgray}{\vhat{Q}}$}
        \linethickness{3pt}
        \color{clgray}
        \put(25,-5){\vector(1,0){40}}
      \end{overpic}
    \end{subfigure} 
    \caption{\label{fig:beta-order} Ordering in \bliiro{}}
  \end{subfigure}
  \begin{subfigure}{\textwidth}
    \centering
    \begin{subfigure}{0.6\textwidth}
      \begin{overpic}[height=0.62\textwidth]{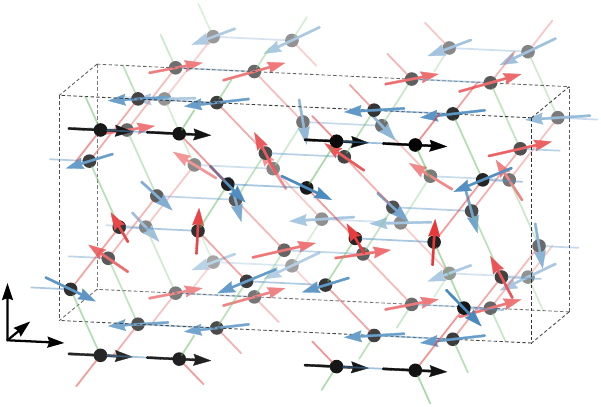}    
        \put(11,10){$\scriptsize \vhat{c}$}
        \put(5,14.5){$\scriptsize \vhat{a}$}
        \put(0.5,21){$\scriptsize \vhat{b}$}
        \put(5,58){$\textcolor{clgray}{\vhat{Q}}$}
        \linethickness{3pt}
        \color{clgray}
        \put(5,53){\vector(1.2,1){10}}
      \end{overpic}
    \end{subfigure}  \hspace{0.25cm}
    \begin{subfigure}{0.3\textwidth}
      \begin{overpic}[width=\textwidth]{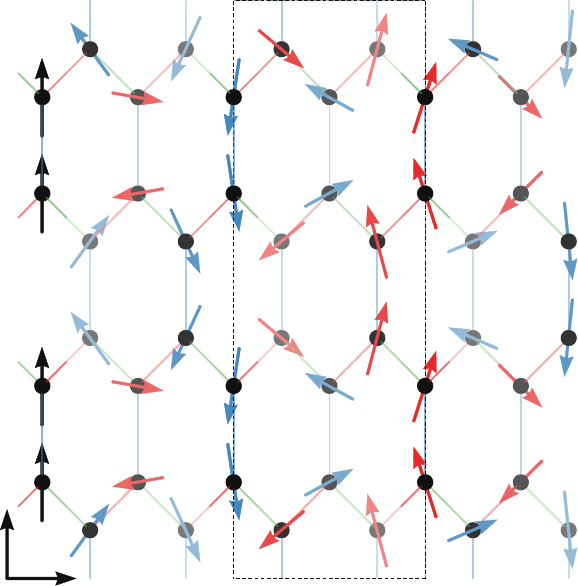}
        \put(14,0){$\scriptsize \vhat{a}$}
        \put(0,14){$\scriptsize \vhat{c}$}
        \put(75,-7){$\textcolor{clgray}{\vhat{Q}}$}
        \linethickness{3pt}
        \color{clgray}
        \put(32,-5){\vector(1,0){40}}
      \end{overpic}
    \end{subfigure} 
    \caption{\label{fig:gamma-order} Ordering in \gliiro{}}
  \end{subfigure}  \caption{\label{fig:orderings} Illustration of the incommensurate,
    non-coplanar, counter-rotating spirals in (a) \bliiro{} \cite{biffin2014unconventional} and (b)
    \gliiro{} \cite{biffin2014noncoplanar}.  For both these materials, the ordering wave-vector is
    aligned to the orthorhombic $\vhat{a}$ direction and three unit cells
    along that direction are depicted.  The non-coplanar nature is
    readily seen in the perspective illustrations on the left while
    the counter-rotating nature of these spiral orders are evident in
    the projected illustration on the right. Color indicates whether
    the $\vhat{b}$ component of the spin is positive (red), negative (blue)
    or zero (black).
    Due to the difference in crystal symmetry of the two lattices, the
    $\vhat{b}$ component of the moments in \gliiro{} transforms differently
    from the $\vhat{a}$ and $\vhat{c}$ components.}
\end{figure}
The hyper- and stripy-honeycomb iridates are built from networks of
edge-shared octahedra much like the 2D honeycomb iridates, but these
networks are intrinsically three-dimensional.  Both of these materials
are insulating, with the magnetic susceptibility indicating a moment
size close to that of a spin-1/2 degree of freedom. As in the case of
\nairo{} and \aliiro{}, this is consistent with a picture of localized
$j=1/2$ moments.  An antiferromagnetic ordering transition is seen at
$\sim 38\K$ in \emph{both} materials through a kink in the magnetic
susceptibility and a peak in the specific heat, though Curie-Weiss
fits demonstrate predominant ferromagnetic exchanges in both
compounds.\cite{takayama2015hyperhoneycomb, modic2014realization} In
the case of \gliiro{}, these exchanges are believed to be anisotropic
as torque magnetometry measurements have revealed a temperature
dependence in the anisotropy of the magnetic
susceptibility.\cite{modic2014realization} Further studies of these
orderings using magnetic resonant x-ray scattering have identified the
magnetic states in both the \bliiro{} and \gliiro{} as incommensurate,
non-coplanar, counter-rotating spirals, as illustrated in
Fig. \ref{fig:orderings}.  Particularly striking is the close
agreement of the incommensurate propagation vectors $\vec{Q} \sim
[0.57,0,0]$ in both compounds.\cite{biffin2014unconventional,
biffin2014noncoplanar} While closely related, these two ordering
patterns have subtle differences: In \bliiro{}, the moments projected
along each of the three orthorhombic directions, $\vhat{a}$,
$\vhat{b}$, and $\vhat{c}$ transform under the same irreducible
representation for a magnetic structure with wave-vector $[q,0,0]$.
In contrast, the $\vhat{a}$ and $\vhat{c}$ components of the spiral
phase of \gliiro{} transform under a different irreducible
representation from that of the $\vhat{b}$ component.  The similarity
of the orderings, ordering temperature, as well as local geometry of
these crystals suggest that a common understanding might be possible.

\subsubsection{Theory}
These experimental discoveries have motivated a number of theoretical
proposals to explain and explore these 3D honeycomb
materials.\cite{lee2014heisenberg, kimchi2014three,
lee2014topological, lee2014order, lee2015theory, kimchi2014unified,
kim2015predominance} Several works attack \bliiro{} and \gliiro{}
directly, starting from a ferromagnetic Kitaev limit and considering
the effects of further perturbations. These include the allowed
nearest-neighbour exchanges discussed in Section \ref{edge-shared}, as
well as those between further neighbours and those induced by the
monoclinic nature of the lattice.  Using these interactions, several
routes have been identified to stabilize an incommensurate spiral
ordering near the ferromagnetic Kitaev limit\cite{lee2015theory,
kimchi2014unified, kim2015predominance}.  Though capable of
stabilizing the ground state, these theories have not been tested
against current experimental findings such as magnetization, torque
magnetometry, and thermodynamic measurements.  Moreover, experiments
such as inelastic scattering that probe low-energy dynamics of these
systems have yet to be conducted.  These comparisons and results
should provide a more comprehensive understanding of the role of
Kitaev physics in the unconventional magnetism encountered in these 3D
honeycomb iridates.

In addition to work related to the experimental magnetic phase,
theoretical work has also focused on the spin liquids that would be
relevant if the Kitaev limit is stabilized on these
lattices.\cite{mandal2009exactly, lee2014heisenberg, kimchi2014three,
nasu2014vaporization,schaffer2015topological, hermanns2015weyl}
Identical to the 2D honeycomb case, the exact solution takes the form
of free Majorana fermions in the presence of a static background $Z_2$
gauge field, and both gapped and gapless phases can be accessed by
tuning the bond anisotropy away from the isotropic limit.  Unlike the
2D case, the lack of mirror symmetry implies that the background flux
generated by the gauge field in the ground state may not be
uniform.\cite{schaffer2015topological} Further, due to the additional
constraints of the three-dimensional geometry, flux excitations of the
$Z_2$ gauge field are restricted to take the form of loops.  These
loops play a key role in the finite temperature properties of the
model, driving a phase transition that separates the low temperature
spin liquid and high temperature
paramagnet\cite{nasu2014vaporization}.  Near the isotropic limit where
the phase is gapless, the spin liquid possesses unusual nodal lines
(co-dimension 2) of gapless Majorana modes.  These nodal lines are
topologically stable and, due to the bulk-boundary correspondence,
induce gapless modes on the surface.\cite{schaffer2015topological} In
the presence of time-reversal symmetry breaking, these line nodes
become topologically protected Weyl points with associated surface
Fermi arcs.\cite{hermanns2015weyl}

Aside from the hyper- and stripy-honeycomb lattices, a number of other
hypothetical trivalent lattices in three-dimensions have also been
considered.  The family of lattices termed the harmonic-honeycombs is
a generalization of the hyper- and stripy-honeycomb lattices.  These
lattices are constructed by exploiting the fact that the hyper- and
stripy-honeycombs can be considered as rows of honeycomb lattices
stacked in an alternating fashion.  By varying the number of complete
honeycomb rows in such a construction, the harmonic-honeycomb series
of lattices is generated.  Each lattice of the series is denoted as
${}^{\mathcal{H}} \langle N \rangle$, where $N$ is an integer that
refers to the number of complete honeycomb rows used in the
construction. \cite{modic2014realization} The hyper- and
stripy-honeycomb lattices are ${}^{\mathcal{H}} \langle 0 \rangle$ and
${}^{\mathcal{H}} \langle 1 \rangle$ respectively in this notation.
Like the hyper- and stripy-honeycomb lattices, the Kitaev spin liquids
in all finite-$N$ harmonic-honeycomb lattices possess topologically
protected line nodes of gapless Majorana fermions and surface gapless
modes.\cite{schaffer2015topological} In contrast, another 3D trivalent
lattice---the hyper-octagon lattice---possesses a gapless phase with a
2D Majorana Fermi surface.\cite{hermanns2014quantum} All of these
theoretical lattices can be embedded in structures containing
edge-shared IrO$_6$ octahedra, hinting at the possibility of realizing
a 3D Kitaev spin liquid in a yet-to-be-discovered material.

\section{Outlook}
\label{sec:outlook}
The wide variety of phenomena observed and predicted in iridates and
related materials has generated immense activity in the field.  Our
objective in this review was not to be exhaustive but rather to
instill a sense of breadth by highlighting a few specific directions
recent studies have pursued.  In particular, we examined the
Ruddlesden-Popper series of iridates and their related
heterostructures where unconventional magnetism, topological band
structures, and superconductivity have been seen or predicted.  We
also examined honeycomb materials, both in two and three dimensions,
where the joined effects of SOC, electron correlation, and lattice
geometry give rise to Kitaev physics and its associated magnetism.
These topics merely scratch the surface of possibilities in the study
of the interplay of SOC and electronic correlations; related research
directions, both theoretical and experimental, further illustrate the
richness of this field.  With this rapid development, however, there
remain several obstacles that if resolved, can substantially enhance
our understanding of SOC effects and propel the community forward.

One issue that has hindered the study of iridate materials is the
large neutron absorption cross-section of Ir.  This property has
restricted neutron scattering experiments to powder samples or large
single crystals, where the former studies lack wave-vector details
while the latter studies may require years of growth technique
refinement.  Although great headway has been achieved using resonant
X-ray scattering as an alternative probe, higher energy resolution and
the ability to perturb the measured system \emph{in situ} would enable
these techniques to be a true alternative to neutron techniques.
Sample quality is also in need of continual refinement in many of the
considered compounds, especially the honeycomb iridates where Na and
Li readily oxidizes, as discussed in Sec. \ref{sec:exp2d}.
Homogeneous solutions of single crystal of \naliiro{} over a wide
range of $x$ (and especially $x=1$) would greatly enhance our
understanding of Kitaev physics and its connection with microscopic
details of materials.  In addition to elemental substitution, other
methods to perturb these systems toward the Kitaev regime or MIT, for
example with applied pressure, could also provide ample insight.  Even
with the well-studied perovskites iridates, there are still
opportunities for major breakthroughs.  Direct evidence of
superconductivity in \sriro{} or other iridium oxides is still
intensely pursued, as is evidence and examples of topological phases.
Aside from strontium iridates, other avenues to realize topological
phases have also been sought.  As mentioned in
Sec. \ref{sec:topological}, engineered phases like superlattices or
thin films of $4d$ or $5d$ TMOs play a special role in this area of
exploration due to their tunability of SOC, correlation, and lattice
distortions.  In this regard, materials like thin film pyrochlore
iridates\cite{hu2012topological, yang2014emergent,
bergholtz2015topology} and epitaxial
perovskites\cite{jang2010electronic, nichols2013tuning,
hirai2015semimetallic, biswas2015persistent} have received
considerable attention.  On the theoretical front, extending our
understanding of topological phases of Sec. \ref{sec:topological}
beyond the non-interacting limit is an important goal.  A large effort
has focused on the consequences of interactions and disorder in Weyl
semi-metals, especially in relation to its
characterization\cite{witczak2014interacting}, transport
properties\cite{burkov2011topological, hosur2012charge}, and
instabilities towards exotic phases\cite{wang2013chiral,
maciejko2014weyl, sekine2014weyl}.  Related is the study of Coulomb
interaction on the quadratic band touching in pyrochlore iridates,
which can induce a quantum critical non-Fermi liquid, a Weyl
semi-metal, or a topological insulator depending on the preserved
symmetries.\cite{moon2013non} These are but some examples where
theoretical predictions of interacting topological phases may find
realizations in iridate materials, and continual search for these
systems will be a central focus in the years coming.

In addition to addressing outstanding issues, the discovery of new
compounds has also driven promising lines of investigation.  These new
materials provide a wonderful arena both to test theoretical
predictions and to discover new phenomena related to strong SOC.  A
recent example is the Mott-insulating Ba$_3$IrTi$_2$O$_9$ that does
not order magnetically.\cite{dey2012spin} It has been argued to
realize the HK model on a triangular lattice, leading to frustration
in both geometry \emph{and} exchanges, which can lead to a variety of
exotic phases.\cite{khaliullin2005orbital,becker2015spin} Another
recently synthesized compound thought to host the HK model is the
honeycomb Li$_2$RhO$_3$, which is closely related to the 2D honeycomb
materials discussed in Sec. \ref{kitaev-two-dimensions}.  This $4d^5$
material does not magnetically order but shows spin-glass behavior
that may be related to disorder or stacking faults.  With this
reported rhodate, alluring ideas like the synthesis of 3D honeycomb
rhodates or isoelectronic substitution of honeycomb materials will
allow the tuning between $4d$ and $5d$ materials to tilt the balance
between SOC, crystal field, and correlation effects.

Intimately related to lattice geometry is how the IrO$_6$ octahedra
are inter-connected; the manipulation of this connectivity has also
been an interesting approach to generate new compounds.  Our review
has focused on corner- and edge-shared octahedra, but there are a
variety of compounds with face-sharing octahedra where SOC plays an
important role in the behavior of the material.\cite{cao2007partial,
terzic2015coexisting} The Mott insulating and magnetically ordered
double perovskites La$_2$ZnIrO$_6$ and La$_2$MgIrO$_6$ serve as yet
another example\cite{cao2013magnetism, cook2015magnetism}: the IrO$_6$
octahedra in these double perovskites are spatially separated because
they do not share corners, edges, nor faces.  The resulting reduction
in orbital overlaps between nearest neighbor $5d$ Ir orbitals leads to
a decrease in bandwidth or, equivalently, a relative enhancement of
SOC and correlation effects.  By further spatially isolating the
IrO$_6$ octahedra, the validity of the localized picture of $j=1/2$
orbitals can be probed.  The magnetically ordered compound
Ca$_4$IrO$_6$ is one such example.\cite{cao2007partial, calder2014j}
The large separation between octahedra in this compound and near-ideal
IrO$_6$ geometry has led to one of the first detections of
distortion-free $j=1/2$ orbitals in this compound.\cite{calder2014j}
The compound Sr$_x$La$_{11-x}$Ir$_4$O$_{24}$ has also been used to
test the $j=1/2$ states but offers an additional twist: with the
tuning of $1 < x < 5$, the valence of the iridium ions can be changed
between Ir$^{4+}$ ($d^5$) and Ir$^{5+}$ ($d^4$).\cite{phelan2015new}
In the strong SOC limit, the $d^4$ filling yields a non-magnetic state
since the atomic $j=3/2$ states are completely occupied and the
$j=1/2$ orbitals remain unfilled.  Measurements of the magnetic
properties of this series of materials indeed show a transition
between the $j=1/2$ and non-magnetic state, indicating that strong SOC
is essential in the description of spatially isolated IrO$_6$
octahedra.  However, this strong SOC limit may not apply to other
compounds with the $d^4$ configuration: examples include the layered
perovskite Ca$_2$RuO$_4$ \cite{mizokawa2001spin}, the honeycomb
materials $A_2$RuO$_3$ ($A$=Li, Na) \cite{wang2014lattice}, and yet
another variety of double perovskites, Sr$_2$YIrO$_6$
\cite{cao2014novel} and La$_2M$RuO$_6$ ($M$=Mg and Zn)
\cite{dass2004ruthenium}. Magnetic transitions have been observed in
these compounds and the origin of such magnetism is under active
investigation.  Proposals such as Van Vleck-type Mott insulators
\cite{khaliullin2013excitonic}, strong non-cubic crystal field effects
\cite{cao2014novel}, and competition between super-exchange and SOC
\cite{meetei2015novel} have been investigated in the context of these
magnetic $d^4$ systems.

Beyond integer fillings, systems with partial filling have also
received attention and the multi-orbital nature of these transition
metal compounds may play an elevated role.  In the recently
synthesized Ba$_5$AlIr$_2$O$_{11}$, dimers of face-shared IrO$_6$
octahedra have an average valence of Ir$^{4.5+}$, yet the material is
a Mott insulator.\cite{terzic2015coexisting} In the thiospinel
CuIr$_2$S$_4$, where sulphur plays the role of oxygen in the
octahedral cages, the average valence is
Ir$^{3.5+}$.\cite{kojima2014magnetic} Several recent studies have
shown that SOC, which have been neglected in earlier
works,\cite{khomskii2005orbitally} may be important in the newly
observed low temperature paramagnetic state.\cite{kojima2014magnetic}
Its cousin, CuIr$_2$Se$_4$, is a metal, but when doped with Pt, shows
evidence of superconductivity: \cite{luo2013superconductivity} an
exciting opportunity considering the local structural similarities
between IrSe$_6$ and the IrO$_6$ octahedra discussed throughout this
article.  Another example is the hyper-kagome, Na$_3$Ir$_3$O$_8$,
which has an average valence of Ir$^{4.33+}$ and thus can be
considered as a 1/3-doped version of the spin-liquid candidate
Na$_4$Ir$_3$O$_8$ mentioned briefly in
Sec. \ref{sec:kitaev-three-dimensions}.\cite{takayama2014spin} Unlike
the Mott-insulating Na$_4$Ir$_3$O$_8$, this new material is
semi-metallic, which was argued to stem from the presence of strong
SOC and distortion-induced molecular orbitals. \cite{takayama2014spin}
With a combination of its chiral crystal structure and frustrated
lattice, it was suggested that this semi-metallic material may harbour
a unique platform to study non-trivial electronic transport due to
topological effects arising from strong SOC.\cite{takayama2014spin}

As we have seen in these brief examples, the study of the combined
effects of strong SOC and correlations has grown to encompass a wide
range of materials and phenomena.  This breadth is an indication of
the rapid development of this field, which boosts the prospects of
discovering novel and exciting physics.  With the eventual advancement
in experimental, numerical, and theoretical techniques, many obstacles
currently faced can be tackled, and perhaps even resolved, thus paving
way for new and exciting research directions in the near future.

\begin{acknowledgements}
  We thank G. K. Khaliullin, Y. J. Kim, Y. B. Kim, B. J. Kim,
P. Gegenwart, J. Matsuno and R. Coldea for useful comments and
discussions. This work is supported by NSERC of Canada (JR, EL, and
HYK).
\end{acknowledgements}

\bibliography{draft}

\end{document}